# Autonomous Multi-objective Alloy Design through Simulation-guided Optimization

*Penghui Yang, Chendong Zhao, Bijun Tang*, Zhonghan Zhang, Xinrun Wang*, Yanchen Deng, Xuyu Dong, Yuhao Lu, Jianguo Huang, Yixuan Li, Yushan Xiao, Cuntai Guan, Zheng Liu*, Bo An**

P. Yang, Dr. Z. Zhang, Dr. Y. Deng, Dr. Y. Lu, J. Huang, Y. Li, Prof. C. Guan, Prof. B. An
College of Computing and Data Science, Nanyang Technological University, Singapore 639798, Singapore
E-mail: boan@ntu.edu.sg

Dr. C. Zhao, Dr. B. Tang, X. Dong, Dr. Y. Xiao, Prof. Z. Liu
School of Materials Science and Engineering, Nanyang Technological University, Singapore 639798, Singapore
E-mail: bjtang@ntu.edu.sg; z.liu@ntu.edu.sg

Prof. X. Wang
School of Computing and Information Systems, Singapore Management University, Singapore 188065, Singapore
E-mail: xrwang@smu.edu.sg

X. Dong
School of Materials Science and Engineering, Shanghai Jiao Tong University, Shanghai 200240, China

Prof. Z. Liu
CINTRA CNRS/NTU/THALES, UMI 3288, Research Techno Plaza, 50 Nanyang Drive, Border X Block, Level 6, Singapore 637553, Singapore
Institute for Functional Intelligent Materials, National University of Singapore, Singapore

Funding: This work was supported by the National Research Foundation Singapore and DSO National Laboratories under the AI Singapore Programme (AISG Award No: AISG2-GC-2023-009-1B). This work was supported in part by the AI2050 initiative at Schmidt Sciences (Grant G-25-68035).

Alloy discovery is constrained by vast compositional spaces, competing objectives, and prohibitive experimental costs. Although simulations and machine learning have each accelerated parts of this process, unifying scientific knowledge, scalable search, and experimental confirmation into a data-efficient workflow remains challenging. Here, we present AutoMAT, a hierarchical autonomous framework spanning ideation to experimental validation. Integrating large language models, automated CALPHAD simulations, residual-learning-based correction, and AI-guided optimization, AutoMAT translates design targets into candidate alloys, refines compositions through closed-loop computational search, and validates results experimentally without hand-curated datasets. Targeting lightweight, high-strength alloys, AutoMAT identifies a titanium alloy 8.1% less dense and 13.0% stronger than the aerospace benchmark Ti-185, achieving the highest specific strength among benchmarked systems. In a second case, AutoMAT discovers a high-entropy alloy with 28.2% higher yield strength than the baseline while preserving high ductility. AutoMAT compresses alloy discovery from years to weeks, establishing a generalizable route toward autonomous materials design.

Penghui Yang and Chendong Zhao contributed equally to this work.

## 1. Introduction

Alloy design underpins key advances in aerospace, automotive, and biomedical sectors[1,2]. From turbine blades to biomedical implants, their broad tunability in mechanical, thermal, and corrosion-resistant properties make them indispensable across industries. To meet escalating performance demands (*e.g.*, higher strength-to-weight ratios and thermal stability), modern alloy design has evolved from systems centered on a single principal element to increasingly complex, multi-component compositions, exemplified by high-strength steels and high-entropy alloys (HEAs)[3,4]. However, this growing compositional complexity introduces a major bottleneck: the design space expands combinatorially, reaching up to $10^{50}$ possible compositions even with a modest number of elements (*e.g.*, 20 elements)[5–7]. Navigating this vast landscape using traditional approaches, which are rooted in experts' intuition and trial-and-error experimentation, is not only time- and cost-intensive, but also inherently inefficient and unsustainable for modern innovation demands[8].

A range of computational methods have been developed to address the limitations of purely experimental alloy discovery, but most still face trade-offs among speed, accuracy, generalizability, and cost. Physics-based simulations such as Computer Coupling of Phase Diagrams and Thermochemistry (CALPHAD) and density functional theory (DFT) can provide physically grounded and interpretable predictions, but are computationally expensive and typically require substantial manual effort, limiting throughput and scalability[9–11]. Moreover, CALPHAD property models employed under standardized, automation-compatible settings may omit certain strengthening mechanisms (e.g., ordering or nanoscale precipitation), introducing systematic biases that compound when predictions are used to guide compositional optimization. Alternatively, machine learning (ML) approaches enable rapid screening across large compositional spaces, but many require large curated datasets and may generalize poorly beyond the training domain[12–14]. The latest developed large language models (LLMs) offer a general framework for knowledge retrieval across domains and can embed physics priors from their internal knowledge. However, they are not inherently built for direct property prediction and alloy design[15]. A unified, scalable solution that combines automation, physical fidelity, and data efficiency has remained out of reach.

Here, we introduce AutoMAT, a hierarchical and autonomous framework for alloy discovery that integrates LLMs, automated CALPHAD simulations, data-driven correction, AI-driven search, and experimental validation into a unified design loop. AutoMAT spans the entire alloy design pipeline, from ideation to simulation and optimization, offering a more scalable, interpretable, and efficient alternative to conventional methods[16,17]. The framework comprises three modular layers. The *Ideation Layer* uses LLMs to extract and propose alloy candidates from literature and handbooks based on user-defined targets. By embedding physics priors from established models and scientific literature, it delivers structured, physically consistent suggestions within minutes. The *Simulation Layer* automates CALPHAD-based thermodynamic modeling and incorporates a data-driven correction module that cross-references CALPHAD predictions against handbook data extracted by the *Ideation Layer*, automatically detecting and compensating for systematic property-model biases with minimal manual intervention. Coupled with AI-guided search, this layer optimizes compositions using orders-of-magnitude fewer evaluations. The final *Validation Layer* physically performs synthesis and characterization of top-ranked candidates, providing critical feedback on real-world performance. As illustrated in **Figure 1**a, AutoMAT integrates the complementary strengths of each method: LLMs for intelligent and data-efficient knowledge retrieval; CALPHAD for thermodynamic accuracy and interpretability; AI-driven search for high-throughput time efficiency and automation. Evaluation criteria for each method are detailed in Section S1 (Supporting Information).

To demonstrate AutoMAT's effectiveness and generalizability, we present two case studies with distinct design goals: the first targets a lightweight, high-strength alloy to show AutoMAT's capability in multi-objective optimization[11]; the second explores a vastly larger design space, identifying a low-cost HEA with high yield strength for potential structural applications. Together, these examples illustrate how AutoMAT accelerates alloy discovery by integrating language understanding, physics-based reasoning, and experimental verification into a hierarchical, autonomous loop.

## 2. System Overview

To operationalize the autonomous and scalable alloy design workflow, we developed AutoMAT as a hierarchical system comprising three modular layers: *Ideation*, *Simulation*, and *Validation* (Figure 1b). Each layer addresses a distinct stage of the discovery process, collectively translating user-defined property targets into experimentally verified alloy compositions through an integrated and largely automated workflow.

The *Ideation Layer* initiates the process by leveraging LLMs, such as ChatGPT and Claude, to propose candidate alloy systems aligned with user-defined property requirements, including high yield strength, low density, or cost constraints (Figure 1c). The first task is alloy system selection. The LLM evaluates which alloy system can satisfy the specific criteria. If no candidate satisfies all targets simultaneously, it recommends the system that best fulfills most objectives, effectively narrowing the compositional design space. Once an alloy system is identified, the LLM extracts candidate compositions from relevant literature, with a preference for alloy handbooks over review articles. Handbooks are priorities because they contain structured, text-based data that is easier for the LLM to parse, whereas reviews often present data in figures or tables, making extraction more difficult. This step outputs scientifically grounded and interpretable suggestions within minutes, significantly accelerating the early-stage ideation phase.

A cornerstone of our methodology is the *Simulation Layer*, which automates CALPHAD thermodynamic calculations within an iterative, AI-guided search loop through a dedicated API (Figure 1d). This transforms CALPHAD from a manually operated modeling tool into a dynamic, on-the-fly component of the discovery pipeline. Starting from the candidate compositions proposed by the *Ideation Layer*, the *Simulation Layer* performs an iterative neighborhood search to refine alloy compositions toward user-defined objectives. At each iteration, elemental ratios are adjusted locally around the current candidate, first using coarse-grained steps to rapidly identify feasible regions and then using finer refinements to improve promising solutions:

$$x_{t+1} = \arg\max_{x' \in \mathcal{N}(x_t)} f(x'), \text{ where } \mathcal{N}(x) = \{x' | x' \text{ } is \text{ } a \text{ } neighbor \text{ } of \text{ } x\}.$$

CALPHAD serves as the backbone of this process, providing physically grounded and interpretable thermodynamic predictions. After acquiring phase volume fractions via Sheil solidification[18], the yield strength can be estimated using the property module (Figure S1, Supporting Information). Compared to surrogate ML models that predict yield strength directly from elemental composition, this physics-informed approach offers more reliable guidance for iterative optimization [19]. The resulting predictions are used to rank candidate alloys and direct the search toward compositions that better satisfy the target property profile. To support automation, simulation conditions are standardized: grain size is fixed at 100 µm, which is the default grain size in the software; only intrinsic strength, grain boundary strengthening, and solid solution strengthening are considered in composition-to-property prediction [20]; and phases with volume fraction less than 2% are excluded. These simplifications improve throughput and consistency across large-scale screening, while parallelized execution enables evaluation of thousands of compositions per day, vastly outpacing traditional expert-driven workflows. The search is run for a fixed number of iterations, even after property targets are met, to allow identification of superior candidates at minimal additional computational cost.

Because the standardized automation settings cannot fully capture all factors governing experimental strength, including omitted strengthening mechanisms and discrepancies introduced during alloy synthesis, processing, and testing, the framework further incorporates a data-driven correction step based on handbook data already retrieved by the *Ideation Layer* (Figure 1d). Specifically, the same CALPHAD pipeline is applied to alloys extracted during ideation, and the predicted values are compared with their reported experimental counterparts. The resulting residuals are then used to train a correction model as a function of alloy composition. During inference, this learned residual is added to the CALPHAD prediction to generate a corrected yield strength for downstream ranking. Notably, this calibration is executed automatically within the pipeline, improving predictive reliability while preserving the scalability and interpretability of the simulation-guided search.

The *Validation Layer* serves as the final stage of the platform, closing the loop by experimentally synthesizing and characterizing the top-ranked candidate alloys to verify computational predictions. This study focuses on as-cast alloys as they reflect the intrinsic mechanical response governed by composition, solid solution strengthening, and grain boundary effects. These unprocessed states serve as a consistent basis for evaluating and optimizing alloy performance. Using arc melting and solidification, followed by tensile testing and microstructural analysis, this layer evaluates real-world properties such as yield strength, density, and phase morphology. Although this step operates on a longer timescale, typically requiring several weeks, the layer provides the essential ground truth needed to confirm the computational predictions and assess their real-world viability. Importantly, the experimental results can be looped back to earlier stages to refine search heuristics and improve the LLM's future recommendations, establishing a closed-loop system for continuous learning and iterative improvement.

By integrating the *Ideation*, *Simulation*, and *Validation Layers* into a cohesive, hierarchical workflow, AutoMAT efficiently translates high-level user requirements into physically validated material solutions with minimal human intervention. Through the combination of rapid candidate generation, automated computational refinement, and experimental confirmation, it enables high-throughput, autonomous discovery of novel alloy compositions tailored to diverse performance objectives.

## 3. Case Study: Design of a Low-Density, High-Strength Alloy

To demonstrate AutoMAT's practical utility and generalizability, we first applied it to a real-world design challenge: developing an alloy that simultaneously achieves low density and high yield strength. The design targets included a yield strength of approximately 850 MPa and a density below 4.36 g/cm$^3$, with the goal of outperforming the state-of-the-art lightweight alloy $Ti_{75.25}Al_{20}Cr_{4.75}$, a benchmark material which is lighter and stronger than most available superelastic alloys, including magnesium-based alloys, nickel-titanium alloys (nitinols) and other titanium-based alloys[21]. Additional constraints were imposed to enhance practical feasibility, including the use of at least four elements and the exclusion of high-cost elements. This multi-objective task reflects critical industrial needs in aerospace and transportation and serves as a stringent test of AutoMAT's capacity to handle competing property requirements, system selection, and design-space refinement.

### 3.1. Literature-Guided Alloy Family Selection and Candidate Generation

The process began in the *Ideation Layer* with alloy system identification (**Figure 2**a). In this step, user-defined targets were translated into a prompt and passed to the LLM to evaluate which alloy system could meet both the high yield strength and low-density requirements. Based on its prior knowledge of lightweight structural alloys, the LLM recommended titanium alloys because they offer relatively low density and together with high mechanical strength.

This early-stage recommendation helped avoid unproductive computational exploration in an unsuitable design space. With titanium alloys selected, the *Ideation Layer* proceeded to handbook-guided candidate selection, where the system accessed a relevant titanium alloy handbook (Figure 2b)[22]. To ensure relevance and feasibility, the LLM was instructed to identify candidate alloys with at least four elements, consistent with multi-principal element design principles, while filtering out compositions containing prohibitively expensive elements. It then processed potential candidates through requirements verification (Figure 2c) and composition filtering (Figure 2d). By parsing textual descriptions, the LLM extracted detailed data on mechanical properties and compositions. Among qualified options, the LLM finally selected $Ti_{86.0}Al_{1.0}V_{8}Fe_{5.0}$ (wt.%), corresponding to $Ti_{86.3}Al_{1.8}V_{7.6}Fe_{4.3}$ (at%), hereafter termed Ti-185, as the most suitable starting point. Ti-185 is a well-documented low-density, high-strength titanium alloy[22]. However, its estimated density (4.70 g/cm$^3$) exceeded the design threshold ($<$ 4.36 g/cm$^3$), necessitating further refinement.

This ideation stage was completed rapidly. Using the GPT-4o API, the LLM identified the alloy system, analyzed the handbook, and selected Ti-185 as the starting candidate within minutes at a computational cost of less than US$1. In contrast, a comparable manual literature search and screening process would likely require several hours or days. These results illustrate the value of the *Ideation Layer* in providing a fast and systematic starting point for downstream simulation-guided optimization. Step-by-step screenshots of the workflow are provided in Figure S2 (Supporting Information).

### 3.2. Compositional Refinement through AI-Guided CALPHAD Optimization

Following the *Ideation Layer*, the initial candidate, Ti-185, was passed to the *Simulation Layer* for further optimization. Given the candidate's relatively high density, the primary objective was to reduce density while maintaining a high yield strength approaching 850 MPa. To achieve this, the *Simulation Layer* initiated an AI-driven iterative neighborhood search, which efficiently explored the compositional space, initially around Ti-185 and subsequently around the best compositions identified in each iteration (**Figure 3**a). At each iteration, candidate compositions were evaluated using CALPHAD-based thermodynamic and property calculations, together with the data-driven correction module that automatically calibrates CALPHAD-estimated yield strengths against handbook-derived records (Figure 3b).

The AI-driven search generated compositional variations by perturbing the current composition, using mole fraction differences to define distances between candidates. The adaptivity of the search strategy was especially crucial here for two reasons. First, there was a significant mismatch between the initial candidate's high density and the desired target range, necessitating aggressive exploration. Second, the candidate's very low aluminum content constrained early compositional changes primarily to increasing the aluminum fraction. This limitation created a sparsely populated region in the compositional space where larger perturbations were both necessary and feasible, justifying the use of coarser step sizes in the initial search phase. Because the system needs to balance two competing objectives, *i.e.*, maximizing yield strength while minimizing density, it becomes challenging to directly compare candidate compositions. To address this, we introduced a score function based on the specific strength, which serves as a single-objective proxy to guide the search. Specifically, we defined a score function as $f = \frac{yield\ strength\ (MPa)}{e^{density\ (g/cm^3)}}$, as shown in Figure 3c. The exponential function of density was empirically chosen to strongly penalize heavier compositions, thereby biasing the search toward alloys with lower density while preserving mechanical strength.

Accordingly, the search began with a coarse-grained approach, employing a step size of 0.5 and a compositional search range of ±10 mol%, with a focus on increasing aluminum content to rapidly shift the design toward lower-density regions. In this stage, the system identified $Ti_{86.0}Al_{10.0}V_{3.5}Fe_{0.5}$, which achieved a substantial density reduction relative to the initial low-Al candidate while preserving high strength. Once promising compositions emerged, the algorithm transitioned to a fine-tuning mode, applying smaller step sizes of 0.2 and a narrowed search range of ±2 mol% for local optimization around the best-performing compositions. After five refinement iterations (detailed in Table S1, Supporting Information), AutoMAT identified an optimal composition, $Ti_{81.4}Al_{16.8}V_{1.6}Fe_{0.2}$ (at%, hereafter termed Ti-185-V), with a predicted density of 4.355 g/cm$^3$ and a predicted yield strength of 771.90 MPa, yielding the highest score in the search trajectory (Figure 3c).

With multi-threaded execution, the *Simulation Layer* evaluated over 1,000 compositions per day. In a broader screening involving over 43,000 potential compositions, the system successfully reduced the candidate pool to 3,161 for detailed evaluation. At an estimated manual rate of 100 CALPHAD evaluations per day, such a task would require approximately two years, whereas AutoMAT completed the screening in less than one week. These results highlight the ability of the Simulation Layer to efficiently navigate a high-dimensional compositional space under competing design objectives. By combining CALPHAD-based evaluation, data-driven correction, and adaptive AI-guided search, AutoMAT identified a low-density, high-strength candidate that provided a strong computational basis for subsequent experimental validation.

**3.3. Experimental Validation of Optimized Alloy Composition.**

Consistent with our focus on as-cast alloys to evaluate intrinsic compositional effects without post-processing, the predicted alloy compositions, Ti-185-V and $Ti_{82.8}Al_{13.0}V_{4.0}Fe_{0.2}$ (at%, hereafter termed Ti-185-III), were experimentally synthesized and validated in the as-cast state using arc melting method. The resulting ingot morphology is shown in Figure 4a. The measured compositions closely matched the recommended values (Table S2, Supporting Information). Tensile testing confirmed that the optimized alloys achieved the targeted combination of low density and high strength (**Figure 4**b, Table S3, Supporting Information). In particular, the synthesized Ti-185-V exhibited a low density of 4.32 g/cm$^3$, satisfying the design requirement of < 4.36 g/cm$^3$, while maintaining a high yield strength of 940 MPa and an impressive specific strength of 243 × 10$^3$ Pa m$^3$ per kg at room temperature. As shown in Figure 4c and Table S4 (Supporting Information), this yield strength exceeds those of deformed Al alloys, deformed Mg alloys, as-cast HEAs, and is comparable to both additive-manufactured titanium alloys and the benchmark $Ti_{75.25}Al_{20}Cr_{4.75}$ shape-memory alloy[21,23–36]. Notably, it exhibits the highest specific strength among all compared systems, underscoring its strong potential for demanding structural applications such as aerospace and space engineering (Figure S3, Supporting Information).

Microstructural characterization revealed several desirable features. The X-ray diffraction (XRD) identified the existence of both α and β phases (Figure S4, Supporting Information). Electron backscatter diffraction (EBSD) and backscattered-electron (BSE) imaging (Figure 4d) showed an ultrafine, lath-like α microstructure interspersed with nanometer-scale phases (phase map in Figure S5, Supporting Information). High-angle annular dark-field scanning transmission electron microscopy (HAADF-STEM) with accompanying energy-dispersive X-ray spectroscopy (EDS) maps (Figure 4e) indicated that these precipitates are Fe- and V-rich (elemental profiles in Figure S6, Supporting Information). A dark-field TEM image acquired using a (010)-type superlattice reflection (Figure S7, Supporting Information) resolved the spatial distribution of bright nano-sized β-phase within the α matrix. Table S5 (Supporting Information) lists the corresponding chemical compositions. HRTEM and the associated fast-Fourier-transform (FFT) patterns verified a fully coherent interface between the α matrix and the nano-sized β phase (Figure 4f). This microstructure, ultrafine lath-like α phases with nanoscale β phases, is known to enhance strength in titanium alloys[37]. Moreover, HAADF-STEM, EDS, and high-resolution TEM (HRTEM) confirmed the presence of the ordered $\alpha_2$ phase within the α matrix, together with slight compositional deviations (Figure S8, Supporting Information). For comparison, theoretical solidification behavior for the present titanium alloys and characterization of the as-cast Ti-185-III alloy are provided in Section S2 (Figure S9 to S12, Table S6 and S7, Supporting Information).

As the aerospace and automotive industries increasingly prioritize fuel efficiency and maneuverability[38,39], lightweight structural materials with superior mechanical properties have become increasingly critical. Ti-185, a widely used commercial titanium alloy, is valued for its exceptional strength-to-weight ratio[22]. In our study, the alloy designed by AutoMAT achieves an 8.1% reduction in density compared to Ti-185 (4.32 g/cm$^3$ vs. 4.70 g/cm$^3$), while increasing yield strength (832 MPa vs. 940 MPa, a 13.0% increase). This is notable given the limited density reduction potential of conventional titanium alloys, which typically range from 4.43 to 4.48 g/cm$^3$ [40, 41]. AutoMAT's ability to overcome the density-strength trade-off highlights its potential to redefine the boundaries of lightweight alloy design. Moreover, In aviation, reducing alloy density offers clear benefits in fuel efficiency, $CO_2$ emissions reduction, and cost savings[42, 43]. Although a reduction in ductility was observed (3.7% vs. 6.9% for Ti-185), this trade-off is acceptable in the context of maximizing specific strength. Together, these results demonstrate that AutoMAT can identify alloy compositions that successfully navigate a challenging density–strength trade-off and translate simulation-guided optimization into experimentally validated materials performance.

**4. Generalizability of AutoMAT**

To further evaluate AutoMAT's scalability and effectiveness across alloy systems of increasing complexity, we applied it to a second design task: discovering a HEA optimized for higher yield strength. This task posed a more demanding challenge due to the high-dimensional design space inherent in multi-component alloy systems, where the number of possible compositions expanded from about 43,000 to over 200,000. Using the AlCoCrFeNi alloy system[44], a widely studied HEA family with promising mechanical properties[45], the *Ideation Layer* proposed Al0.5-Co-Cr-Fe-Ni as the starting point. The *Simulation Layer* then refined the composition through five iterations of AI-driven search ultimately identifying $Al_{14.5}Co_{27.0}Cr_{21.5}Fe_{13.0}Ni_{24.0}$ (at%) (Figure 3d). Detailed optimization procedures are available in Section S3 (Figure S13, Table S8, Supporting Information). This process reduced the candidate pool from over 200,000 possible compositions to fewer than 6,000 candidates for detailed evaluation, compressing simulation time from an estimated ten years of manual effort to only two weeks. Experimental validation confirmed a 28.2% improvement in yield strength over the baseline while maintaining high ductility. This combination is especially valuable for structural applications requiring both strength and ductility. This case study highlights AutoMAT's scalability and effectiveness in high-dimensional compositional spaces, with detailed characterization results available in Section S4 (Figure S14 and S15, Table S9 and S10, Supporting Information) and detailed cost analysis in Figure S16 (Supporting Information).

**5. Discussion**

AutoMAT's strength lies in its modular and generalizable architecture. The *Ideation Layer*, powered by LLMs, interprets unstructured scientific knowledge to suggest candidate systems based on user-defined targets. This makes the framework domain-agnostic and easily extensible beyond metals to ceramics, polymers, catalysts, biomaterials, any field with sufficiently rich textual data. The *Simulation Layer* supports plug-and-play integration with different physics engines. While CALPHAD was employed here for its efficiency and thermodynamic rigor, other simulation methods such as DFT, molecular dynamics, or finite element analysis can be incorporated for different material classes or target properties.

The data-driven correction module exemplifies a broader quality-assurance strategy enabled by the framework's layered design: during pipeline construction, cross-referencing CALPHAD predictions against handbook data extracted by the Ideation Layer revealed systematic deviations in yield-strength estimates, which were automatically compensated by training a lightweight regression model on the residuals. Because this self-calibration mechanism relies solely on the literature data already harvested by the *Ideation Layer*, it generalizes naturally to any simulation engine for which comparable reference values exist, requiring no additional experimental input or domain expertise. This inter-layer synergy, in which handbook data simultaneously informs initial composition selection and calibrates simulation fidelity, illustrates how each module in AutoMAT reinforces the others, positioning the framework as a robust and extensible platform for autonomous materials innovation.

Taken together, the two case studies validate AutoMAT's ability to address multi-objective optimization tasks and navigate high-dimensional design spaces, and demonstrate that its benefits are not confined to a single alloy family. More broadly, AutoMAT embodies a forward-looking paradigm for materials discovery, in which LLMs provide literature-grounded priors, AI-driven search methods efficiently explore vast and complex design spaces, and physically grounded simulations rigorously evaluate candidate solutions. This synergistic integration transforms unstructured domain knowledge into a navigable design landscape, enabling discovery that is both fast and physically meaningful.

**6. Conclusion**

In summary, this work introduces AutoMAT, a hierarchical and autonomous framework that spans the entire alloy discovery pipeline from ideation to experimental validation. Across two case studies, AutoMAT delivers substantial acceleration and performance gains: in the Ti-Al-V-Fe system, it discovers a lightweight titanium alloy that is 8.1% less dense than the aerospace-grade Ti-185 while raising yield strength from 832 MPa to 940 MPa (a 13.0% increase), achieving the highest specific strength among all benchmarked systems; in a high-entropy alloy case, it identifies a composition with 28.2% higher yield strength than the baseline while preserving high ductility.

Looking forward, the generalization capability of LLMs makes AutoMAT continuously adaptive, serving as a dynamic scientific assistant that can synthesize evolving knowledge. AI-driven search pushes boundaries toward previously uncharted regions of materials space, while simulations ensure predictive fidelity before advancing to experimental testing. Together, these elements form the foundation for fully agentic workflows that can propose, assess, and optimize materials with minimal human intervention, paving the way for faster, more data-efficient, and scalable discovery across all material classes.

## 7. Methods

### 7.1. LLM-driven Ideation

OpenAI's GPT-4o model (accessed via API) was utilized for all natural language processing and knowledge extraction tasks within the *Ideation Layer*. GPT-4o was selected for its advanced capabilities in natural language understanding, complex reasoning, and information extraction from extensive textual corpora at the time of the study, which were deemed critical for effectively navigating and interpreting materials science literature and specialized handbooks.

User-defined property targets, such as desired yield strength (*e.g.*, approaching 850 MPa), maximum density (*e.g.*, < 4.36 g/cm$^3$), and elemental cost or exclusion constraints, were formulated into structured prompts to guide GPT-4o. All prompt templates were set up in advance with blanks in them. When the user inputs the requirements, their requirements will be filled into the blanks to complete the prompt. All prompt templates can be found in Section S5 (Supporting Information).

The system outputs scientifically grounded, structured suggestions in JSON format for downstream processing. This initial stage was highly efficient, generating a viable starting candidate within minutes at a computational cost of less than US$1, a process that would typically require several hours of manual literature search. More detailed cost comparison can be found in Fig S18.

### 7.2. Automated CALPHAD Simulations

All thermodynamic calculations and properties predictions were conducted using Thermo-Calc software (version 2024b). Automation of these computations was achieved through the TCPython package interfacing with the server. The TCHEA7 thermodynamic database underpinned all calculations, supplying the Gibbs energy descriptions for constituent phases essential for determining equilibrium states.

The initial step in all calculations involved computing phase volume fractions. For this, the Scheil module was utilized to generate Scheil solidification curves. These curves represent the phase formation during solidification by sampling numerous points between a fraction solid of 0 and 1, thereby identifying the corresponding phases at each interval. The system incorporates a process to identify new phases as the temperature drops and filter noise from the Scheil curve, yielding accurate final phase volumes. Apart from the composition itself, temperature is the only critical parameter, which was set to 3000 °C to ensure complete melting of the alloys under investigation.

The specific property calculations were performed with established phase volumes. Densities, for both individual phases and the overall alloy, were derived from the molar volumes and compositions of the equilibrium phases. The yield strength of candidate alloys was predicted via the property model calculation module, considering intrinsic strength, solid solution strengthening, and grain boundary strengthening. To enable full automation and ensure consistency, several assumptions were standardized. A grain size of 100 μm was specified for the grain boundary strengthening contribution. Furthermore, phases comprising less than 2% of the total volume fraction were excluded from calculations, treated as computational noise to simplify the analysis.

**7.3. Data-Driven Yield Strength Correction**

While the standardized CALPHAD settings described above enable full automation, they omit certain strengthening contributions (e.g., precipitation hardening) that can be significant in real alloys. To quantify and correct this discrepancy automatically, the framework incorporates a data-driven correction module as a built-in quality-assurance step within the *Simulation Layer*. This module is applied in both case studies presented in this work; the titanium alloy case is described here as an illustrative example. Because the *Ideation Layer* identified Ti-185, a Ti-Al-V-Fe alloy, as the starting composition, the pipeline automatically retrieved all handbook alloys whose constituent elements are a subset of {Ti, Al, V, Fe} from the titanium alloy handbook. This element-based filtering ensures that no extraneous alloying elements enter the training set, thereby minimizing the number of input features and keeping the resulting correction model as simple as possible. After further filtering for data quality, nine alloy entries were retained as training samples. Compositions were converted from weight percent to atomic percent to match the simulation input format, and CALPHAD yield strength was computed under the identical conditions used in the *Simulation Layer* (i.e., Scheil solidification, 100 μm grain size, solid solution and grain boundary strengthening only).

A correction model was trained to predict the residual $\Delta = YS_{experimental} - YS_{CALPHAD}$ using alloy composition (Al, V, and Fe in at%) as input features. Given the small number of both training samples and input features, a first-degree Ridge regression with regularization parameter $\alpha = 10.0$ was adopted to mitigate overfitting. At inference, the corrected yield strength is computed as $YS_{corrected} = YS_{CALPHAD} + \Delta_{predicted}$. The corrected yield strengths for the candidate paths in both the titanium alloy and HEA case studies are reported in Table S1 and S8 (Supporting Information), respectively. The regression fit on the training set and the out-of-sample validation against synthesized alloys are detailed in Section S6 and Figure S17 (Supporting Information). This correction module operates as a fully automated post-processing step within the *Simulation Layer*, with no manual parameter tuning or domain-expert input required.

**7.4. AI-driven Compositional Search**

An AI-driven iterative neighborhood search was used to refine the LLM-generated candidates and explore the compositional space. The search algorithm iteratively seeks to optimize a target function by exploring neighboring compositions: $x_{t+1} = \arg\max_{x' \in \mathcal{N}(x_t)} f(x')$, where $\mathcal{N}(x) = \{x' | x' \text{ } is \text{ } a \text{ } neighbor \text{ } of \text{ } x\}$.

In Case Study 1, to handle the multi-objective optimization task of balancing low density and high strength, a custom score function was defined as $f = \frac{yield\ strength\ (MPa)}{e^{density\ (g/cm^3)}}$. The exponential penalty on density was empirically chosen to strongly bias the search toward lighter alloys while preserving high strength. The search began with a coarse-grained approach, employing a step size of 0.5 and a compositional search range of ±10 mol%, with a focus on increasing aluminum content to rapidly shift toward lower-density regions. In this stage, the system identified $Ti_{86.0}Al_{10.0}V_{3.5}Fe_{0.5}$, which achieved a substantial density reduction relative to the initial low-Al candidate while preserving high strength. Once promising candidates were identified, the algorithm transitioned to a fine-tuning mode using smaller step sizes of 0.2 and a narrowed search range of ±2 mol% for local optimization around the best-performing compositions. All the search steps are shown below:

- Step 1: $Ti_{86.0}Al_{10.0}V_{3.5}Fe_{0.5}$ → ρ = 4.460 g/cm$^3$, $YS_{CALPHAD}$ = 961.42 MPa, $\Delta YS$ = -184.89 MPa, $YS_{corrected}$ = 776.53 MPa, score = 8.979.

- Step 2: $Ti_{84.2}Al_{11.8}V_{3.8}Fe_{0.2}$ → ρ = 4.437 g/cm$^3$, $YS_{CALPHAD}$ = 952.37 MPa, $\Delta YS$ = -177.55 MPa, $YS_{corrected}$ = 774.82 MPa, score = 9.167.
- Step 3: $Ti_{82.8}Al_{13.0}V_{4.0}Fe_{0.2}$ → ρ = 4.427 g/cm$^3$, $YS_{CALPHAD}$ = 944.88 MPa, $\Delta YS$ = -174.30 MPa, $YS_{corrected}$ = 770.58 MPa, score = 9.209.
- Step 4: $Ti_{82.2}Al_{15.0}V_{2.6}Fe_{0.2}$ → ρ = 4.387 g/cm$^3$, $YS_{CALPHAD}$ = 935.35 MPa, $\Delta YS$ = -163.84 MPa, $YS_{corrected}$ = 771.51 MPa, score = 9.596.
- Final step: $Ti_{81.4}Al_{16.8}V_{1.6}Fe_{0.2}$ → ρ = 4.355 g/cm$^3$, $YS_{CALPHAD}$ = 927.08 MPa, $\Delta YS$ = -155.18 MPa, $YS_{corrected}$ = 771.90 MPa, score = 9.913.

**7.5. Specimen preparation and mechanical tests**

Ingots were prepared by arc-melting the high-purity elements of Ti, Al, V and Fe, Co, Cr, Ni (larger than 99.99 wt.%). The ingots were re-melted at least seven times to ensure chemical homogeneity under high-purity argon atmosphere. The molten alloys were then suction-cast into a water-cooled copper mold with dimensions of diameter of 10 mm and length of 100 mm. To determine their mechanical performances, flat, dog-bone-shaped tensile samples with a gauge length of 15 mm, a width of 2.5 mm and a thickness of 1.5 mm were machined from the alloys by using electrical-discharge machining. Uniaxial tensile tests were conducted using a ETM503C universal testing machine at a strain rate of $1 \times 10^{-3}$ s$^{-1}$ at room temperature, and each condition was tested for at least three times to ensure the repeatability of our testing. The density of the alloys at each condition was measured 5 times via SD-200L densimeter. The chemical composition of each ingot was measured via inductively coupled plasma emission spectrometer (ICP) method.

**7.6. Microstructure observations and crystal structure characterizations**

Phase constitutions were analyzed by MA Xima-7000 X-ray diffractometers (XRD) conducted with Cu Kα radiation in the 2θ range of 30°-90° and a scan speed of 4°/min. Microstructure observations and phase distributions were realized by Zeiss Gemini 300 equipped with Oxford Symmetry S3 electron backscatter diffraction detector. TEM analysis was conducted using a ThermoFisher Talos F200X fitted with a STEM-EDS detector to investigate their crystal structure, elemental compositions, and interfacial features between different phases at 200 kV. The High-resolution TEM (HRTEM) image was acquired on an aberration-corrected JEOL ARM-300F at 300 kV to observe the atomic structure. The TEM specimens were mechanically polished to a thickness of 50 μm and electropolished in a twin-jet polishing machine (Struers TenuPol-5).

**Acknowledgements**

We acknowledge Thermo-Calc for providing the shrinkage calculations. This work was supported by the National Research Foundation Singapore and DSO National Laboratories under the AI Singapore Programme (AISG Award No: AISG2-GC-2023-009-1B). This work was supported in part by the AI2050 initiative at Schmidt Sciences (Grant G-25-68035).

**Conflict of Interest**

The authors declare that they have no known competing financial interests or personal relationships that could have appeared to influence the work reported in this paper.

**Author contributions**

P.Y. and C.Z. contributed equally to this work. B.A., Z.L., X.W. and B.T. conceived and supervised the project. P.Y. designed the AutoMAT framework and developed the algorithms. C.Z. synthesized the samples and performed mechanical and structural characterizations. Z.Z. contributed to the CALPHAD calculations. X.D. contributed to the data-driven correction module. X.W., Y.D. and Y.L. supported the development of the AI search algorithm. Y.L., J.H. and Y.X. assisted with data analysis. P.Y., C.Z., and B.T. co-wrote the manuscript. All authors discussed the results and contributed to the final manuscript.

**Code and Data Availability Statement**

The code used in this study is available at https://github.com/penghui-yang/AutoMAT under the MIT license. The repository contains all scripts necessary to reproduce the results presented in this paper, along with usage instructions and example datasets. Any additional code used for analysis is available from the corresponding author upon reasonable request. The authors declare that all relevant data are included in the paper and its Supplementary Information.

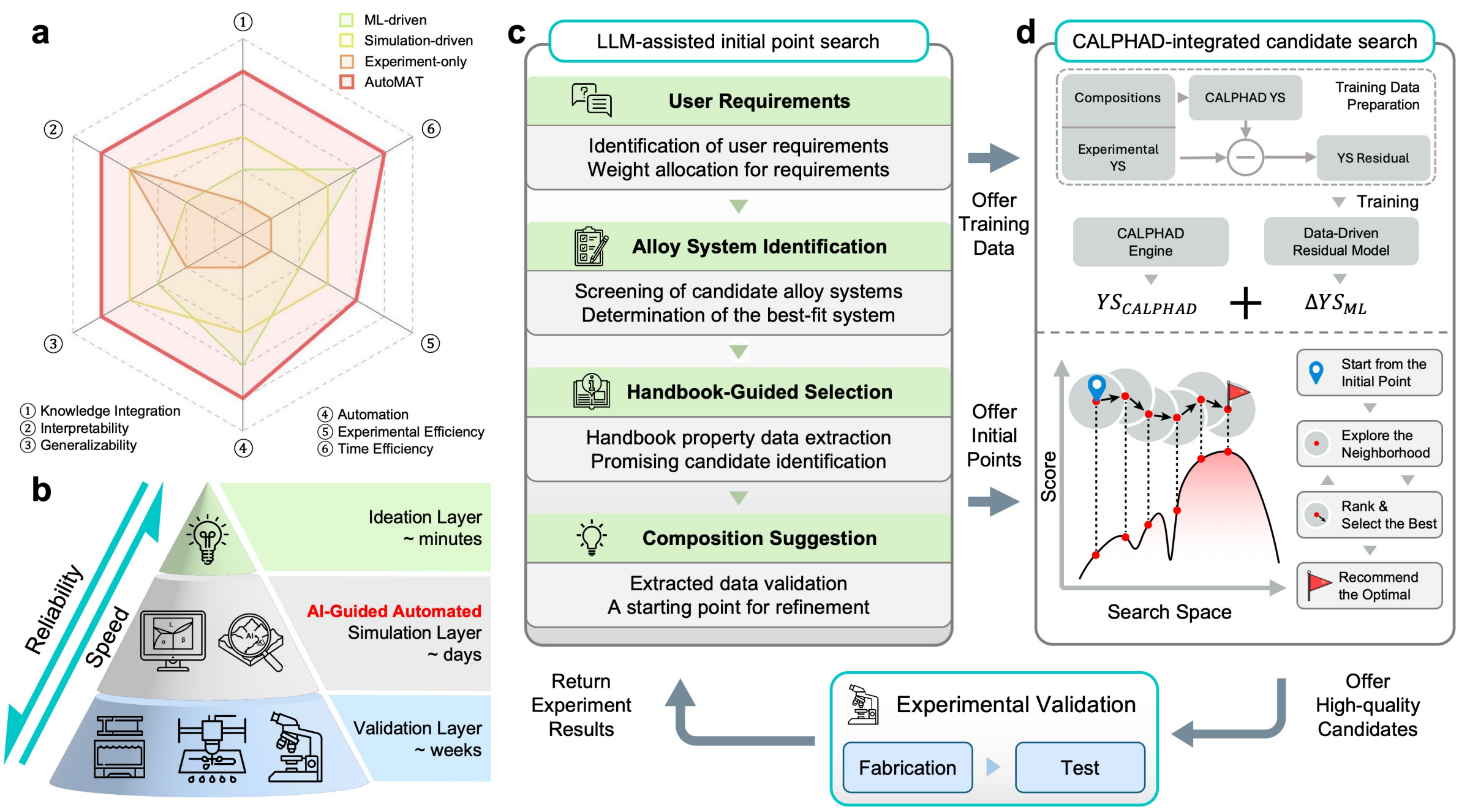

**Figure 1.** AutoMAT framework: architecture and performance comparison. a) Radar chart comparing the performance of AutoMAT with existing methods across six key dimensions: knowledge integration, interpretability, generalizability, automation, experimental efficiency, and time efficiency. Scoring criteria are detailed in Section S1 (Supporting Information). b) Overview of AutoMAT's three-tier structure: the Ideation Layer, the Simulation Layer, and the Validation Layer. c) In the Ideation Layer, LLMs assist in alloy system selection and initial composition identification based on user-defined property targets, handbook data, and known alloy systems. d) The Simulation Layer applies CALPHAD-based modeling to explore compositional neighbourhoods, rank candidates, and recommend optimal compositions for experimental validation. The CALPHAD-based modeling is enhanced by a data-driven model which predicts the residual of the CALPHAD predicted yield strength and the experimental yield strength.

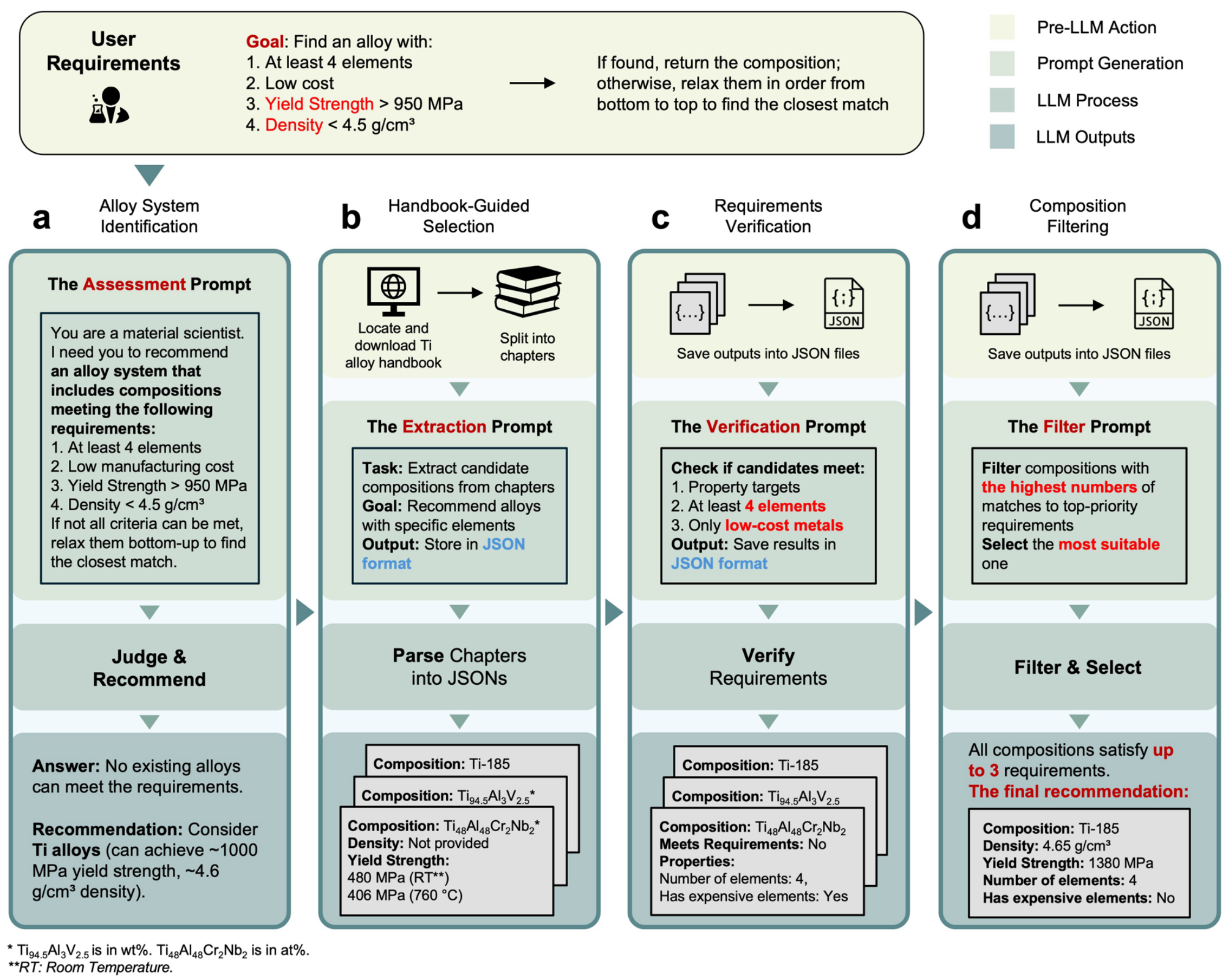

**Figure 1.** Stepwise process of the Ideation Layer in AutoMAT. a) Alloy system identification proposes an alloy system that best matches the user-defined requirements. b) Handbook-guided selection retrieves the alloy handbook for the chosen system and extracts all possible alloy compositions. c) Requirements verification evaluates whether each candidate meets the specified property constraints. d) Composition filtering ranks candidates based on the degree of requirements satisfaction and selects the most promising composition.

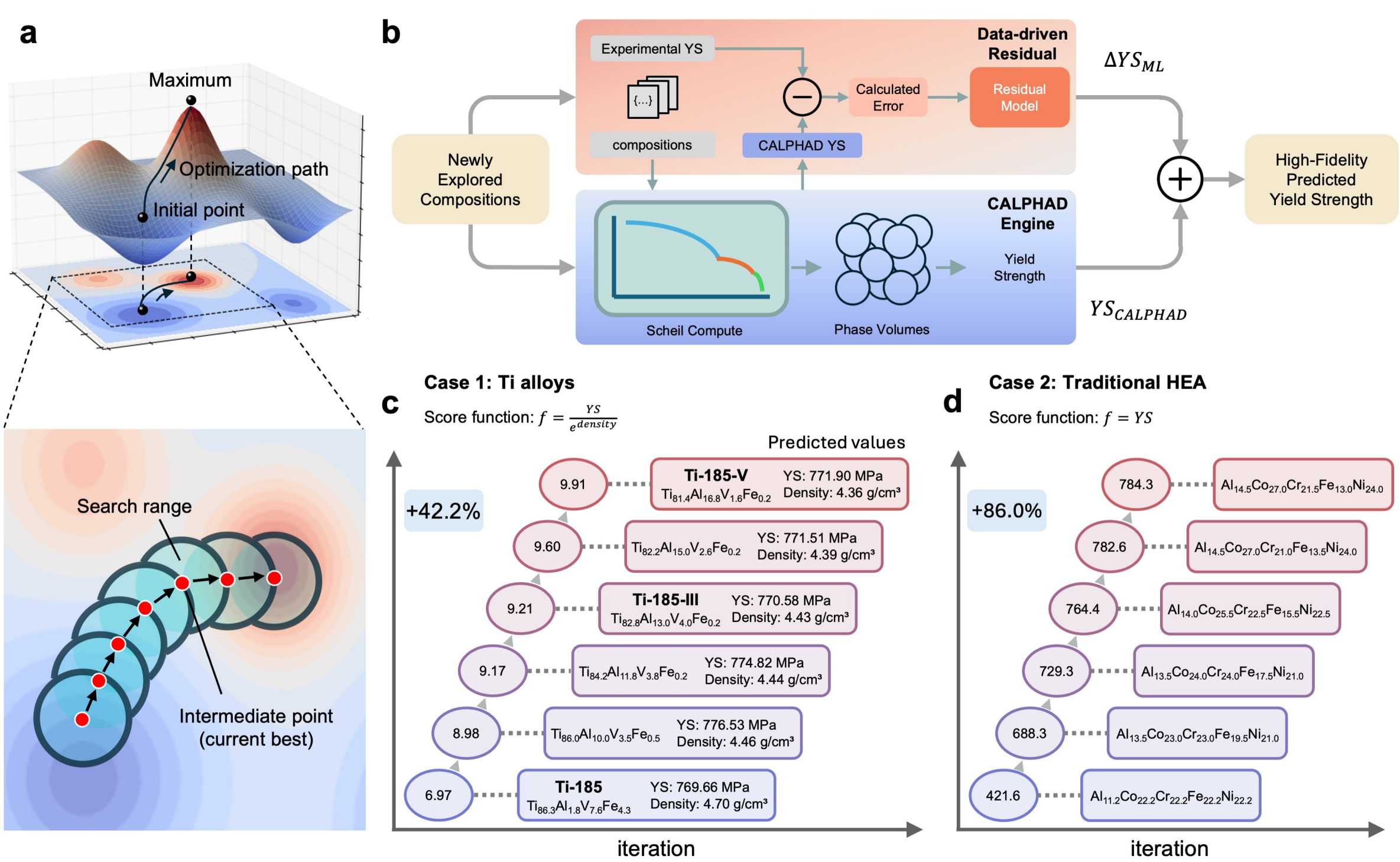

**Figure 2.** AI-guided CALPHAD optimization and search trajectories in AutoMAT. a) AI-driven search explores the compositional design space by iteratively sampling within a fixed search range around the current candidate. b) Schematic of the data-driven yield strength correction module. A regression model, trained on handbook alloys with known experimental yield strengths, learns the systematic residual $\Delta = YS_{experimental} - YS_{CALPHAD}$. At inference, the predicted residual is added to the raw CALPHAD output to produce a calibrated yield strength estimate. This correction is executed automatically within the Simulation Layer as a built-in quality assurance step, requiring no expert intervention. c) Search trajectory for Case 1, showing a 42.2% score improvement as the model iteratively proposes alloys with enhanced yield strength and reduced density. d) Search trajectory for Case 2, achieving a 86.0% improvement through identification of compositions with markedly higher mechanical performance.

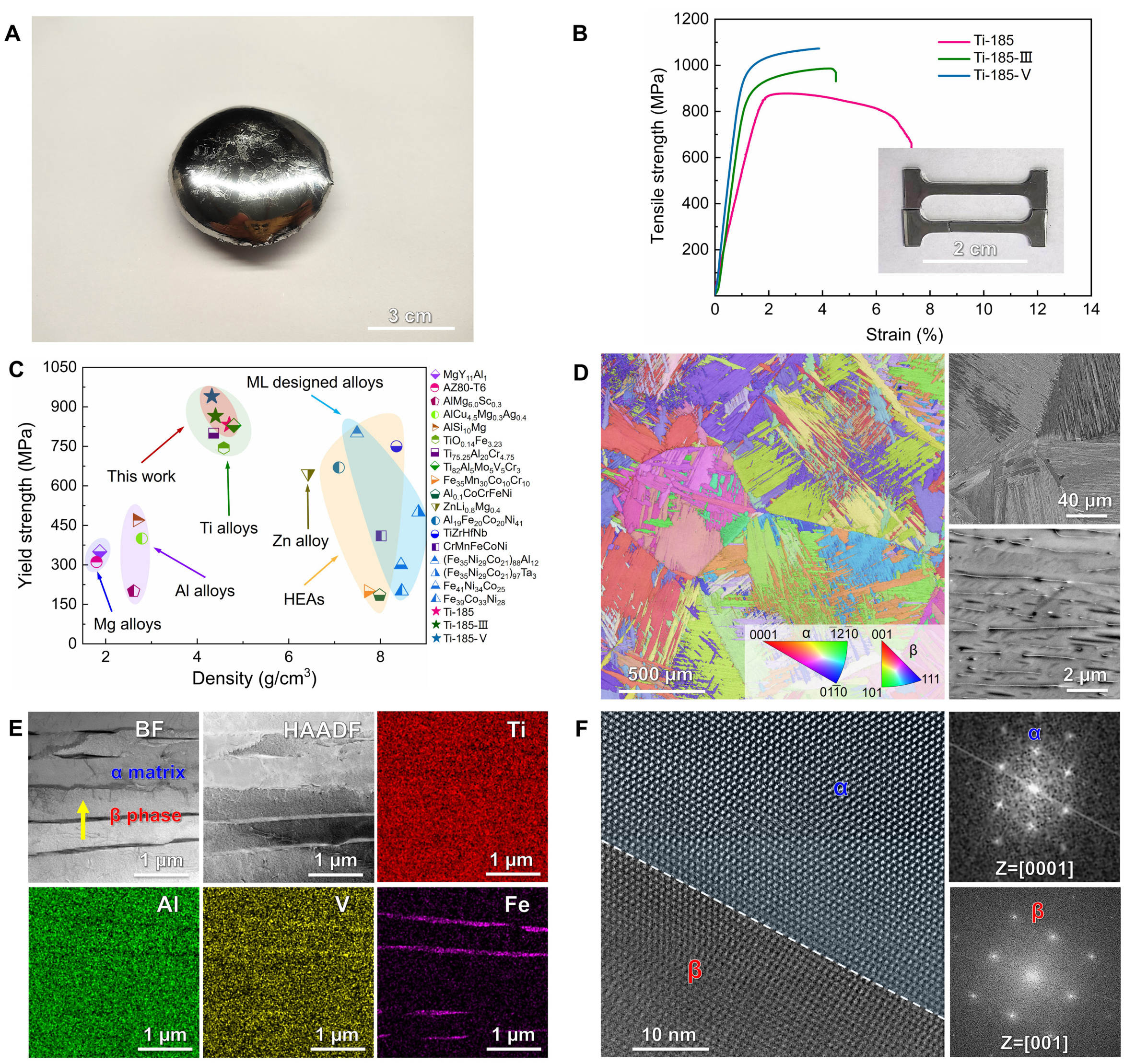

**Figure 3.** Mechanical performance and microstructural characterization of the as-cast titanium alloys. a) The morphology of the ingot. b) Representative engineering stress-strain curves from tensile testing; the inset shows the morphology of the tensile specimens. c) Comparison of yield strength and density between the current study and previously reported alloys[21,23–36]. d) The EBSD inverse pole figure (IPF) color map of the as-cast $Ti_{81.4}Al_{16.8}V_{1.6}Fe_{0.2}$ alloy; right panels display BSE images at different magnifications. e) Bright-field TEM image, HAADF-STEM image, and EDS mappings of the α-β region of the as-cast $Ti_{81.4}Al_{16.8}V_{1.6}Fe_{0.2}$ alloy. f) High-resolution TEM image of the α-β region of the as-cast $Ti_{81.4}Al_{16.8}V_{1.6}Fe_{0.2}$ alloy, with its corresponding FFT images displayed on the right.

## The table of contents entry

AutoMAT, a hierarchical autonomous framework, integrates large language models, automated CALPHAD simulations, residual-learning-based correction, and AI-guided optimization to translate design targets into validated alloys without hand-curated datasets. It discovers a titanium alloy 8.1% lighter and 13.0% stronger than aerospace benchmark Ti-185, and a high-entropy alloy with 28.2% higher yield strength at preserved ductility, compressing discovery from years to weeks.

## Autonomous Multi-objective Alloy Design through Simulation-guided Optimization

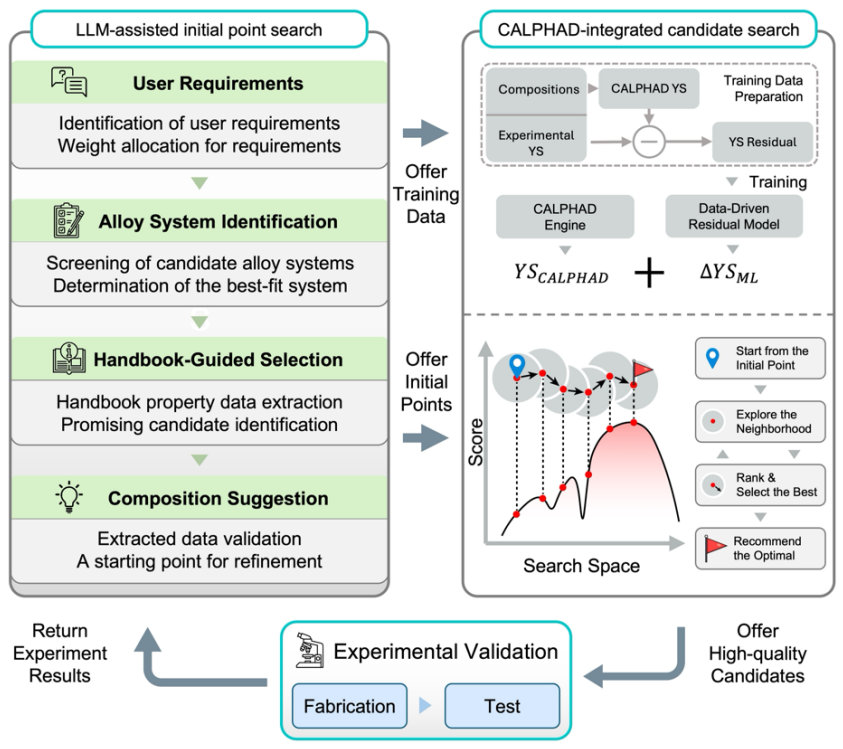

# Supporting Information

**Autonomous Multi-objective Alloy Design through Simulation-guided Optimization**

*Penghui Yang, Chendong Zhao, Bijun Tang*, Zhonghan Zhang, Xinrun Wang*, Yanchen Deng, Xuyu Dong, Yuhao Lu, Jianguo Huang, Yixuan Li, Yushan Xiao, Cuntai Guan, Zheng Liu*, Bo An**

## 1. Scoring rules of the radar chart

- **Knowledge Integration**

  • Definition: Refers to the method's ability to systematically utilize and integrate existing human knowledge, such as scientific literature, manuals, and databases.

  • Scoring Criteria:

  - 1 point: Hardly uses existing knowledge bases; mainly relies on intuition or raw data.
  - 2 points: Indirectly utilizes knowledge through training data; unable to actively query.
  - 3 points: Depends on formalized knowledge bases (e.g., thermodynamic databases).
  - 4 points: Can combine multiple formalized knowledge bases for decision-making.
  - 5 points: Actively and systematically integrates unstructured textual knowledge (literature) and structured data (manuals, databases).

- **Interpretability**

  • Definition: Each decision step and the final outcome provided by the method are clearly traceable and supported by explicit physical or chemical principles.

  • Scoring Criteria:

  - 1 point: Completely "black-box"; decision processes cannot be explained.
  - 2 points: Partially interpretable; critical steps depend on hard-to-understand correlations.

- 3 points: Results interpretable, but the deduction process is indirect and convoluted.
- 4 points: Most decisions based on explicit physical models (e.g., thermodynamics).
- 5 points: Every step, from initial selection to final optimization, is supported by a clear, intuitive physical/chemical logical chain.

- **Generalizability**
  • Definition: The framework's ability and associated cost to transition from one materials discovery problem (e.g., specific alloy systems or properties) to a completely different materials discovery problem. High generalizability indicates universal core logic and low transition cost.
  • Scoring Criteria:
    - 1 point: Highly problem-specific, almost impossible to transfer; relies on implicit, non-transferable expert knowledge.
    - 2 points: Requires substantial time investment to learn; explicit but non-transferable expert knowledge.
    - 3 points: Transferable but heavily dependent on extensive data from the specific domain. Transferring requires expensive new data collection and model training.
    - 4 points: Framework depends on a single type of physical model/database within a specific domain. Transfer depends on the existence or development of similar dedicated computational tools/databases in the new domain.
    - 5 points: Framework is modular; the transfer process is clear, involving only replacement of core computational modules (e.g., replacing CALPHAD with DFT) while maintaining the overall logic.
- **Automation**
  • Definition: The degree to which tasks in the discovery process can be automatically executed by machines or software, reducing human intervention.
  • Scoring Criteria:
    - 1 point: Completely manual; every step requires human operation and decision-making.

- 2 points: Partial computational automation; most tasks still require human intervention.
  - 3 points: Partial computational automation; human intervention required to connect workflow steps.
  - 4 points: Most computational workflows automated; initial parameters and interpretation of results require human input.
  - 5 points: Most steps in the closed loop are automated, with human involvement primarily for supervision.

- **Experimental Efficiency**
  • Definition: The number and complexity of experiments required throughout the discovery process to identify the final candidate alloy.
  • Scoring Criteria:
  - 1 point: Extremely high; requires numerous and complex experimental series.
  - 2 points: Relatively high; requires multiple iterative experiments for training or validation.
  - 3 points: Moderate; validation experiments needed for a set of candidate materials.
  - 4 points: Relatively low; validation needed only for a few (<5) final candidate materials.
  - 5 points: Extremely low; ideally only the final 1-2 optimal candidates require validation.

- **Time Efficiency**
  • Definition: The time required from project initiation to obtaining validated candidate alloys that meet the specified requirements.
  • Scoring Criteria:
  - 1 point: Extremely slow, typically taking several years.
  - 2 points: Slow, taking more than one year.
  - 3 points: Moderate, ranging from several months to one year.
  - 4 points: Relatively fast, usually completed within three months.

- 5 points: Extremely fast, theoretically able to complete a discovery cycle within weeks.

**2. Theoretical solidification behaviors and microstructural characterizations of the as-cast Ti-185 and $Ti_{82.8}Al_{13.0}V_{4.0}Fe_{0.2}$ alloys**

The simulated results indicated that both the theoretical crack susceptibility coefficient (CSC) and volumetric shrinkage of the as-cast titanium alloys exhibited only minor variations, consistent with the excellent casting qualities shown in Figure 4a (Table S6). The XRD results confirmed the coexistence of both α and β phases in both as-cast Ti-185 and $Ti_{82.8}Al_{13.0}V_{4.0}Fe_{0.2}$ alloys (Figure S9). SEM and EBSD results revealed that the $Ti_{82.8}Al_{13.0}V_{4.0}Fe_{0.2}$ alloy possessed an overall lath-like α–β morphology (Figs. S10 and 11). Furthermore, HAADF-STEM and HR-TEM observations identified Fe- and V-rich β precipitates with coherent interfaces to the α phase (Figure S12). The detailed chemical compositions of both α and β phases are shown in Table S7.

**3. Details of the case study on conventional HEAs in the *Simulation Layer***

In the case of conventional HEAs, the initial candidate Al0.5-Co-Cr-Fe-Ni was passed to the *Simulation Layer* for optimization. Given the candidate's relatively low initial predicted yield strength of 532.11 MPa, the primary objective was to maximize this property. The *Simulation Layer* initiated its autonomous pipeline by leveraging CALPHAD-based methodologies for precise thermodynamic and physical property predictions and a data-driven correction module. Its core strategy was an AI-driven iterative neighborhood search, which efficiently explored the compositional space, initially around the starting HEA and subsequently around the best compositions identified in each iteration (Figure S13).

At each iteration, CALPHAD simulations were performed to predict key properties, with a focus on yield strength, which was calculated based on the volume fractions of individual phases predicted by CALPHAD. Then a data-driven correction module was introduced to get a corrected predicted yield strength. Since the system needed to optimize a single objective yield strength, the comparison between candidate compositions was straightforward. Specifically, we defined the score function as $f = Yield\ Strength\ (MPa)$, as shown in Figure 3d. This score function directly guided the search toward alloys with the highest possible mechanical strength. The algorithm deployed a stable step size of 0.5 for local optimization around the best-performing compositions (Table S8). All the search steps are shown below:

- Step 1: $Al_{13.5}Co_{23.0}Cr_{23.0}Fe_{19.5}Ni_{21.0}$ → YS = 688.26 MPa.
- Step 2: $Al_{13.5}Co_{24.0}Cr_{24.0}Fe_{17.5}Ni_{21.0}$ → YS = 729.25 MPa.
- Step 3: $Al_{14.0}Co_{25.5}Cr_{22.5}Fe_{15.5}Ni_{22.5}$ → YS = 764.36 MPa.
- Step 4: $Al_{14.5}Co_{27.0}Cr_{21.0}Fe_{13.5}Ni_{24.0}$ → YS = 782.55 MPa.
- Final step: $Al_{14.5}Co_{27.0}Cr_{21.5}Fe_{13.0}Ni_{24.0}$ → YS = 784.29 MPa.

Over the course of five refinement iterations, the simulation converged on a composition that significantly enhanced the target property. The final composition, $Al_{14.5}Co_{27.0}Cr_{21.5}Fe_{13.0}Ni_{24.0}$, achieved a predicted yield strength of 784.29 MPa, representing a 86.0% improvement over the initial candidate. This case study underscores the *Simulation Layer*'s central role in AutoMAT. By integrating rigorous CALPHAD predictions with adaptive AI-guided search, and applying domain-informed constraints to support automation, AutoMAT efficiently explored the high-dimensional compositional space to achieve a targeted design objective. The results demonstrate AutoMAT's capability to accelerate the discovery of high-performance alloys through efficient, autonomous decision-making.

## 4. Experimental validation of *Simulation Layer* based on AlCoCrFeNi alloy system

The chemical composition of the computationally optimized HEAs was close to their experimental compositions (Table S9). Moreover, the theoretical crack susceptibility coefficient (CSC) was decreased and volumetric shrinkage of the as-cast HEAs varied slightly, indicating an improvement of the casting qualities (Table S6). XRD results showed that the HEAs were composed of both FCC and BCC phases (Figure S14). Tensile test results confirmed the success of AutoMAT's prediction and optimization process of the yield strength (Figure S15). The yield strength increased from 305 MPa to 397 MPa, and the density decreased from 7.33g/cm$^3$ to 7.17g/cm$^3$ (Table S10).

## 5. Prompt and examples of LLM responses in the *Ideation Layer*

**Step 1** Alloy System Identification

**User:**

You are a material scientist. I need you to recommend an alloy system that includes components meeting the following requirements:

1. At least 4 elements

2. Low manufacturing cost
3. Yield Strength > 950 MPa
4. Density < 4.5 g/cm$^3$

If not all criteria can be met, relax them bottom-up to find the closest match.

**Assistant (LLM):** To meet your tough targets, you need lightweight and high-strength systems. Here are some promising directions:

1. Titanium Alloys

Density: ~4.43 g/cm$^3$

Yield Strength: Often exceeds 800 MPa, with certain grades like Ti-6Al-4V reaching yield strengths of 900–1000 MPa.

Composition: Primarily titanium, aluminum (6%), and vanadium (4%) in wt %.

Performance: Commonly used in aerospace and biomedical applications due to their excellent strength-to-weight ratio and corrosion resistance.

2. Aluminum-Lithium Alloys

Density: Typically around 2.6–2.8 g/cm$^3$

Yield Strength: Exceeds 800 MPa in some grades, especially when heat-treated.

Composition: Aluminum, lithium (1–3%), with trace amounts of other elements like copper or magnesium.

Performance: Used in aerospace to reduce weight while maintaining strength, particularly in aircraft fuselage and wings.

3. Magnesium Alloys

Density: Around 1.74–2 g/cm$^3$

Yield Strength: Newer magnesium alloys have been developed with yield strengths exceeding 300 MPa and can reach up to 800 MPa under the right conditions.

Composition: Magnesium combined with elements like aluminum, zinc, and rare earth elements.

Performance: Magnesium alloys are of particular interest for automotive and electronics industries where weight savings are critical.

**Step 2** Handbook-Guided Selection

**User:** I have provided a chapter in a material handbook in PDF format. The chapter contains information about the properties and components of various alloys. I would like you to act as a material scientist and help me analyze the chapters.

1. Your Task:

Retrieve all the components that have the exact information of Yield Strength and Density in the book.

2. Expected Output:

List the recommended alloy components along with their corresponding properties (density, yield strength, etc.). Please retrieve all the results with information of Alloy Component, Density (g/$cm^3$) and Yield Strength (MPa) in a json format.

**Assistant (just one example is shown here):** Sure! Based on the provided content from the PDF, here is the information in JSON format regarding the alloys with their corresponding Density and Yield Strength:

```json
[
  {
    "Alloy Component": "Corona 5",
    "Density (g/cm^3)": 4.54,
    "Yield Strength (MPa)": 825
  },
  {
    "Alloy Component": "Ti-6Al-4V",
    "Density (g/cm^3)": 4.43,
    "Yield Strength (MPa)": 795
  }
]
```

This JSON format lists the relevant alloy components and their properties as you requested.

**Step 3** Requirements Verification

**User:** I have provided a component extracted from a material handbook. The contents contain information about the properties and components of various alloys. I would like you to act as a material scientist and help me analyze the contents.

Your Task:

Evaluate the alloy component based on the primary requirements provided. Your task is to determine if it meets the primary requirements.

Primary Requirements (Must-Have):

- The component should contain four elements or more.

- The component should include low-cost metals such as Al, Fe and exclude any expensive metals such as Zr, Mo.

Expected Output: Output whether it meets the primary requirements. If it does, specify the properties (number of elements and cost-related details). If not, state which requirement(s) it fails to meet.

Format: Please present the result in a clear json format as follows:

```json
{
  "Alloy Component": [Name of Component]
  "Meets Requirements": [Yes/No]
  "Properties":{
    "Number of elements": [Number]
    "Has expensive elements or not": [Yes/No]
  }
}
```

**Assistant (just one example is shown here):**

```json
```

```
{
  "Alloy Component": "Ti-6Al-4V",
  "Meets Requirements": "No",
  "Properties": {
    "Number of elements": 3,
    "Has expensive elements or not": "No"
  }
}
```

Explanation:

- Ti-6Al-4V is a titanium alloy that primarily consists of titanium (Ti), aluminum (Al), and vanadium (V), making it a three-element alloy.

- It does not meet the requirement of containing four or more elements.

- While it includes aluminum, a low-cost metal, it does not contain any explicitly listed expensive metals like zirconium (Zr) or molybdenum (Mo). Therefore, it does not have expensive elements. However, titanium itself is generally considered a relatively expensive material.

**Step 4** Component Filtering

**User:** I have provided some filtered contents from a material handbook. The contents contain information about the properties and components of various alloys. I would like you to act as a material scientist and help me analyze the contents.

1. Your Task:

Recommend the relevant alloy components and their properties that meet my requirements.

2. Property Requirements:

Based on the filtered data, recommend alloy components that meet the following property requirements:

Requirements (Nice-to-Have):

- Have clear and unambiguous components.

- Density smaller than 3.8 (g/cm$^3$).

- Yield Strength larger than 1000 (MPa).

You need to first list all the components that satisfy the primary requirements, and then choose those that meet the requirements.

If no information in the document(s) fully satisfies all the requirements, try to recommend components that satisfies as much requirements as possible.

For requirements that are satisfied by the components, it is encouraged to exceed the requirements so that we can have more space for further modification.

For components that have the same performance on one requirement, prefer those perform better on other requirements.

3. Expected Output:

List the recommended alloy components along with their corresponding properties (density, yield strength, etc.).

Then point out the most recommended one.

4. Format:

Please present the results in a clear table format, with columns for:

- Alloy Component

- Density (g/cm$^3$)

- Yield Strength (MPa)

- Number of elements

- Has expensive elements or not

**Assistant:** To address the property requirements specified, I will begin by analyzing the components within the list provided and check how they align with your stated conditions:

Requirements:

- Density smaller than 3.8 g/cm$^3$

- Yield Strength larger than 1000 MPa

- Clear and unambiguous component

Upon reviewing the retrieved contents, none of the alloy components meets the exact requirement of having a density smaller than 3.8 g/cm$^3$. However, I can identify those components which have the highest potential, exceeding in yield strength and performing best based on the available properties.

Here is the list of alloy components meeting or exceeding the highest possible criteria of yield strength:

| Alloy | Density (g/cm$^3$) | Yield strength (MPa) | Number of elements | Has expensive elements or not |
|---|---|---|---|---|
| Ti-8V-5Fe-1Al | 4.65 | 1380 | 4 | No |
| Ti-10V-2Fe-3Al | 4.65 | 1240 | 4 | No |
| Ti-15V-3Cr-3Al-3Sn | 4.76 | 1165 | 5 | No |
| Ti-6-22-22S Billet | 4.65 | 1089 | 4 | No |
| Ti-15-3 (Aged) | 4.76 | 1025 | 4 | No |

Recommendation:

Since none of the alloys meets the density requirement, we focus instead on yield strength performance. Here are the top performers based on yield strength:

- **Ti-1Al-8V-5Fe** (wt.%): This component provides the highest yield strength of 1380 MPa among the listed alloys, with moderate density and four elements, making it versatile given its absence of expensive elements.

- **Ti-3Al-10V-2Fe** (in wt.%, Yield strength: 1240 MPa): If prioritized by yield strength while maintaining clarity in the component composition, this is a strong candidate.

Based on the yield strength performance, the **most recommended alloy** is **Ti-1Al-8V-5Fe**, due to its highest yield strength and moderately favorable density, aligning with your emphasis on exceeding requirements.

While all these recommendations don't fully meet your original request for density, the listed alloys potentially provide excellent yield strength characteristics.

## 6. Evaluation of the Data-driven Correction Module

The importance of the correction step is best appreciated by examining both its in-sample fit and its out-of-sample predictive behaviour. Figure S17 shows the regression performance on the handbook training set. Because the model is intentionally constrained to a linear form with strong regularization, the in-domain fit appears modest; however, this conservative choice is deliberate, as a more flexible model would risk overfitting the limited training data and producing unreliable extrapolations for unseen compositions.

The practical value of the correction was confirmed in the *Validation Layer* in the Ti-alloy case, where three synthesized titanium alloys, none of which were included in the training set, served as independent test cases. The raw CALPHAD predictions for Ti-185, Ti-185 III, and Ti-185 V were 1023.9, 944.9, and 927.1 MPa, respectively, exhibiting a monotonically decreasing trend along the composition path. After correction, the predicted values became 769.7, 770.6, and 771.9 MPa, reversing the trend to a monotonically increasing one. The experimentally measured yield strengths of 831, 865, and 943 MPa confirmed this increasing trend. Although the corrected predictions underestimate the absolute experimental values, they successfully capture the correct compositional ranking across the alloy series, which is the primary objective of the screening stage. As an automatically derived calibration that requires no domain expertise, the correction model provides a useful complement to CALPHAD-based screening by improving candidate ranking without compromising the scalability or interpretability of the pipeline.

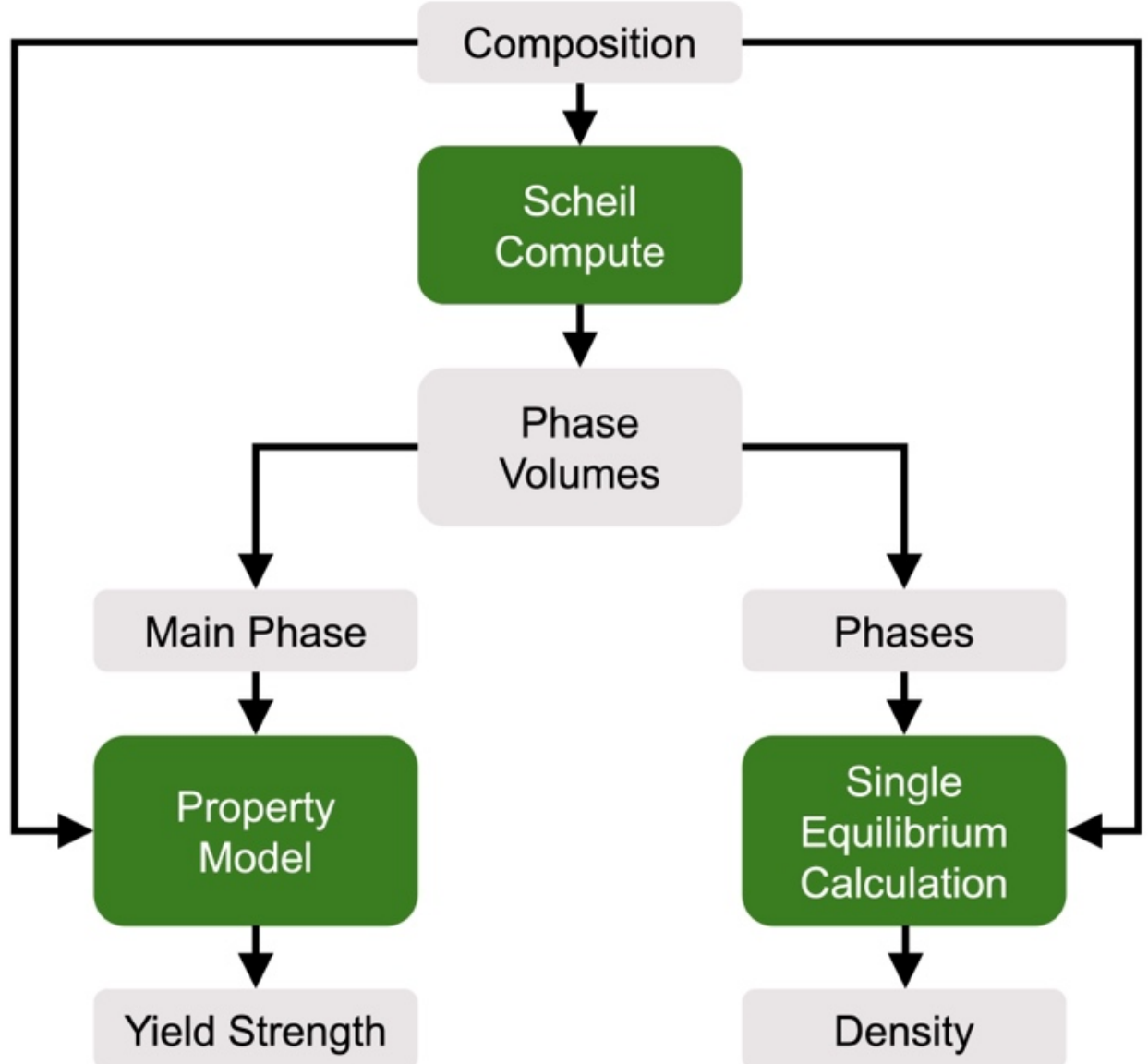

**Figure S1.** Workflow of the CALPHAD-based property prediction within the *Simulation Layer*. For a given composition, Scheil computation simulates non-equilibrium solidification to obtain phase volume fractions. The main phase is identified and fed, together with the composition, into the property model to calculate yield strength. In parallel, the full set of phases is used in a single equilibrium calculation to estimate density.

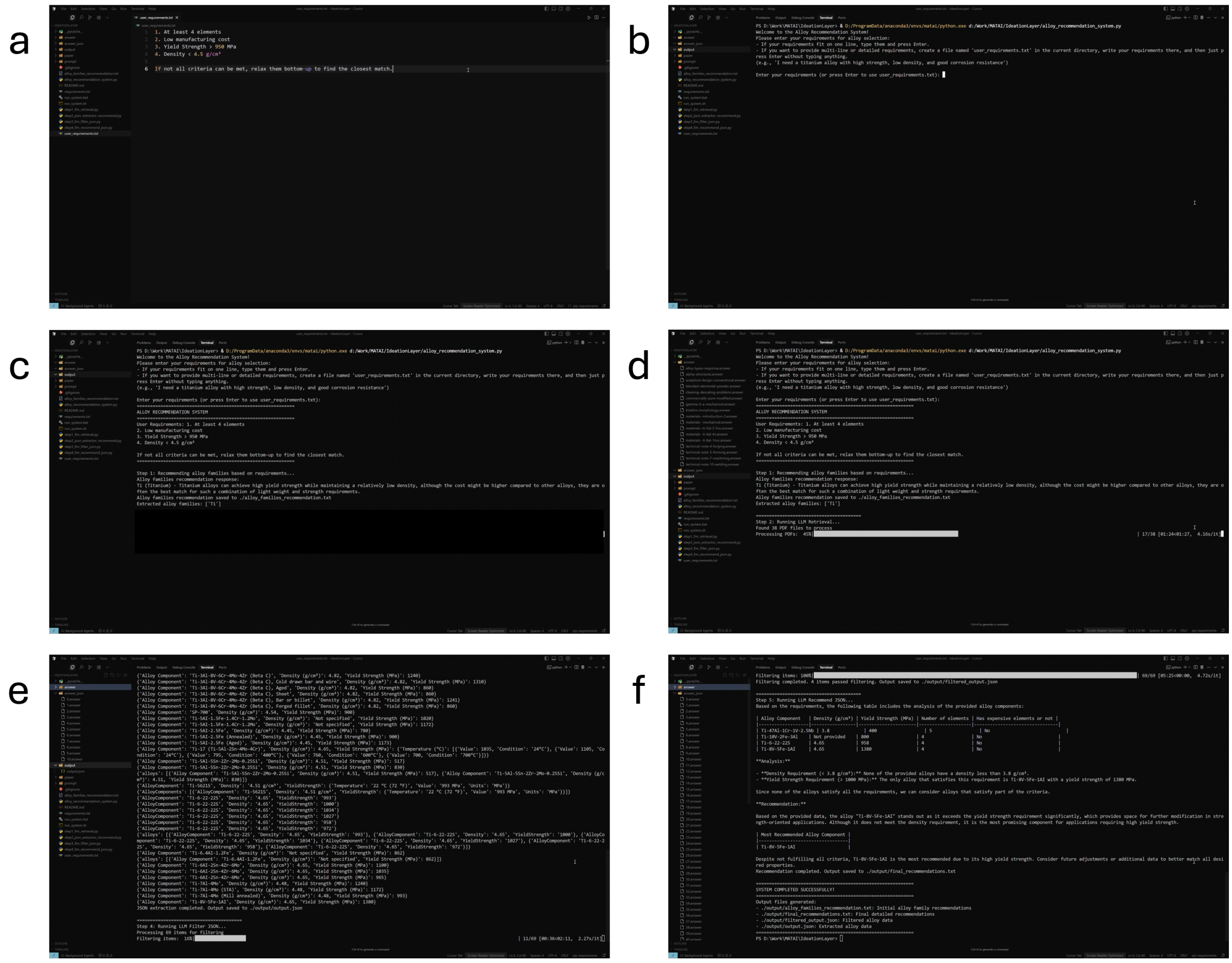

**Figure S2.** Screenshots of the *Ideation Layer*'s workflow. a) Input the user requirements into the file. b) Press Enter when the system asks for the user requirements. c) The *Ideation Layer* recommends the Ti alloy family. d) The *Ideation Layer* retrieves the Ti alloy handbook and extracts all possible alloy components. e) The *Ideation Layer* evaluates whether each candidate meets the specified property constraints. f) The *Ideation Layer* finally selects the most promising composition: Ti-185.

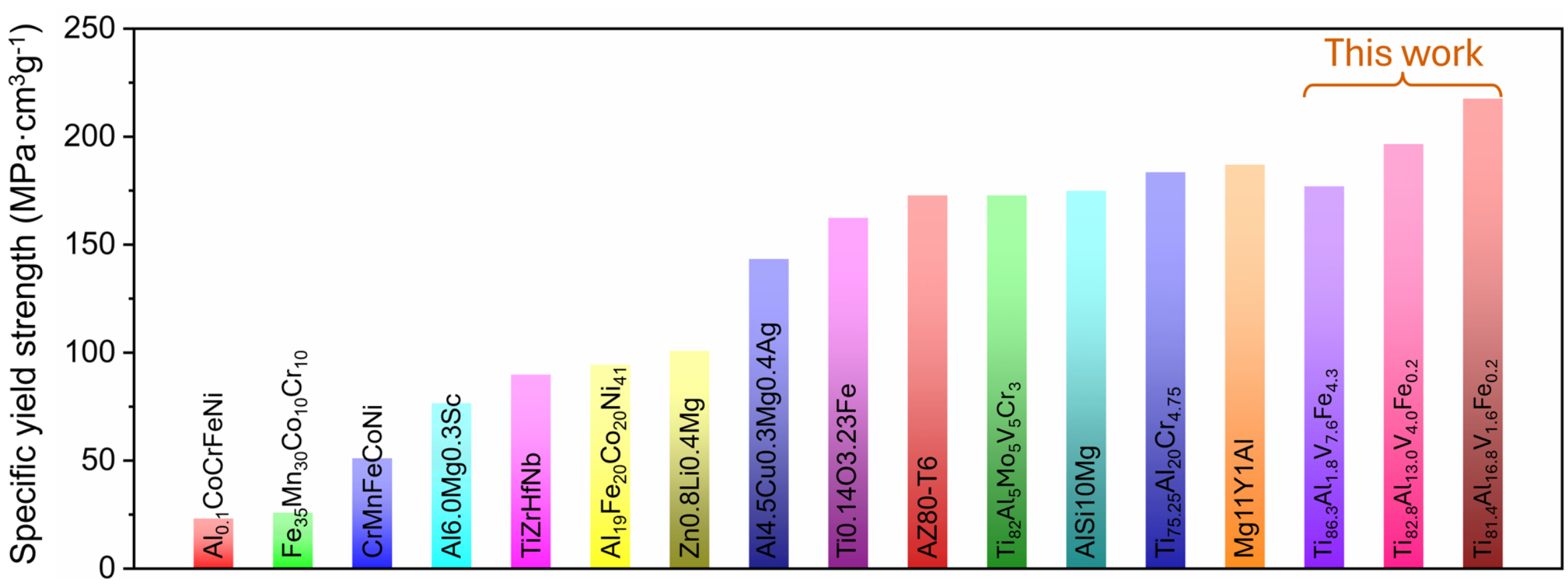

**Figure S3.** Comparison of specific yield strength AutoMAT-designed alloys with state-of-the-art alloys. The $Ti_{81.4}Al_{16.8}V_{1.6}Fe_{0.2}$ alloy exhibits the highest specific yield strength among deformed Al alloys, deformed Mg alloys, as-cast HEAs, and additive-manufactured titanium alloys.

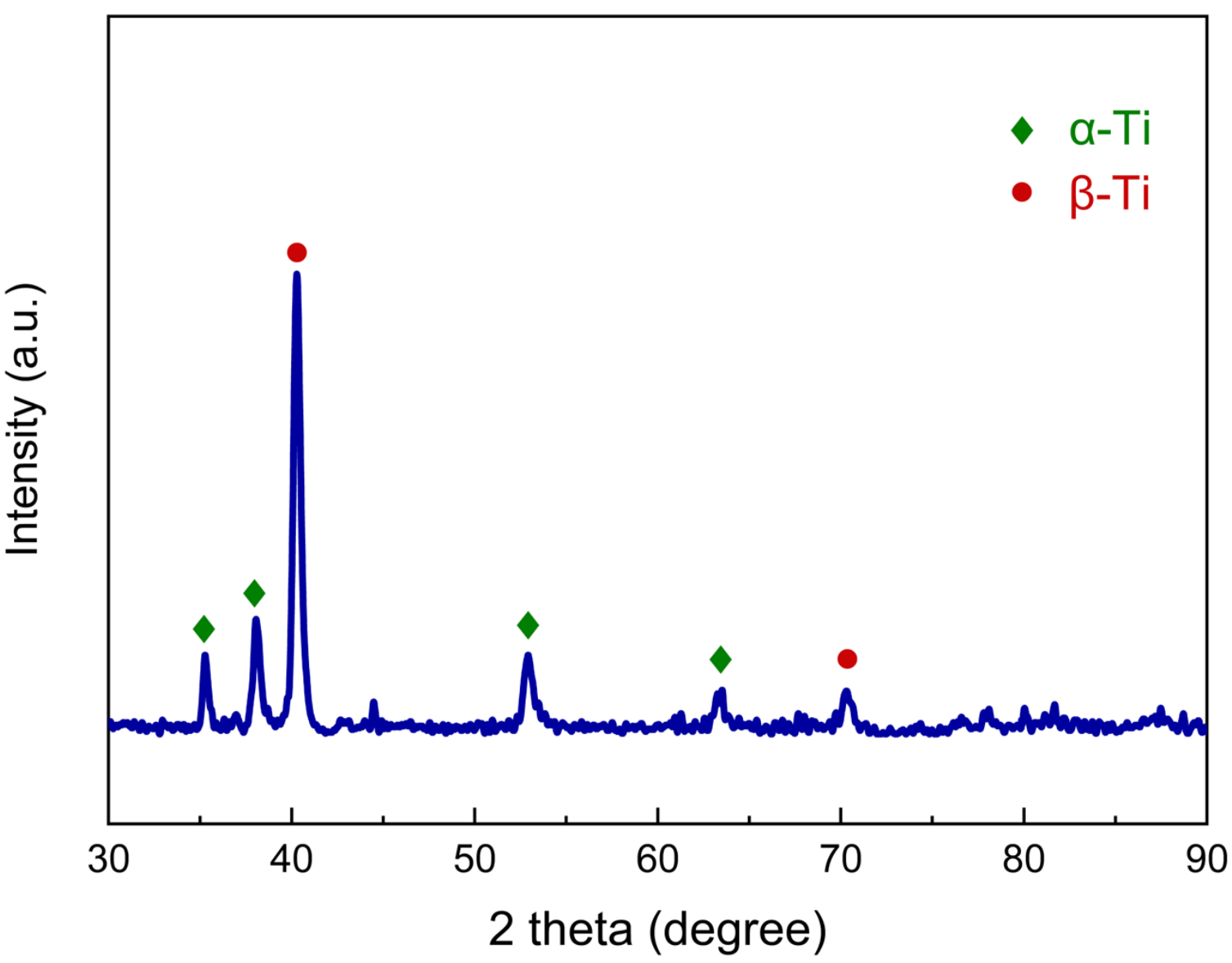

**Figure S4.** XRD pattern of the as-cast $Ti_{81.4}Al_{16.8}V_{1.6}Fe_{0.2}$ alloy. The result shows that the alloy was composed of α phase and β phase.

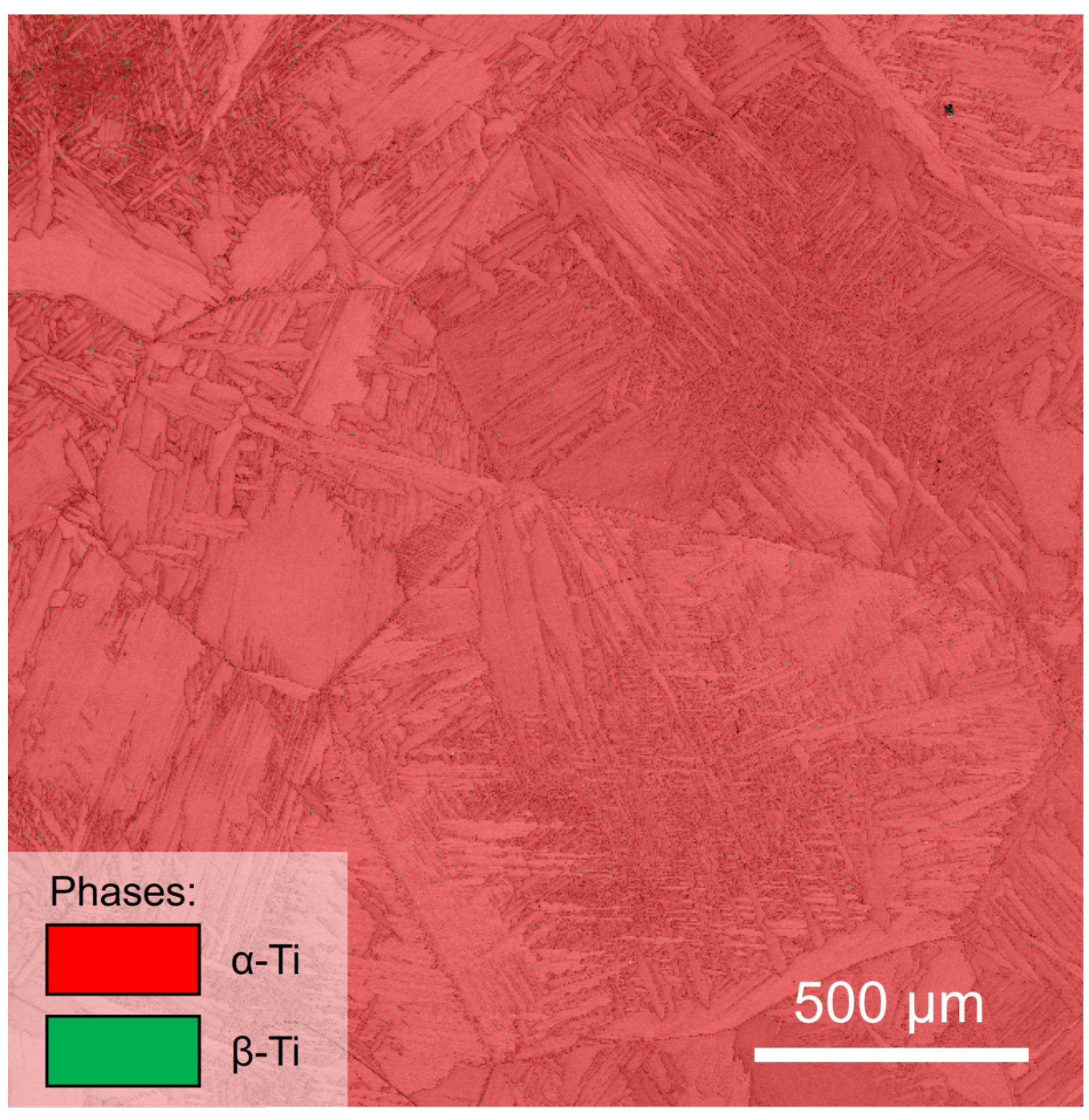

**Figure S5.** The corresponding EBSD phase map of Figure 4d. The result shows that the alloy was mainly composed of the α phase.

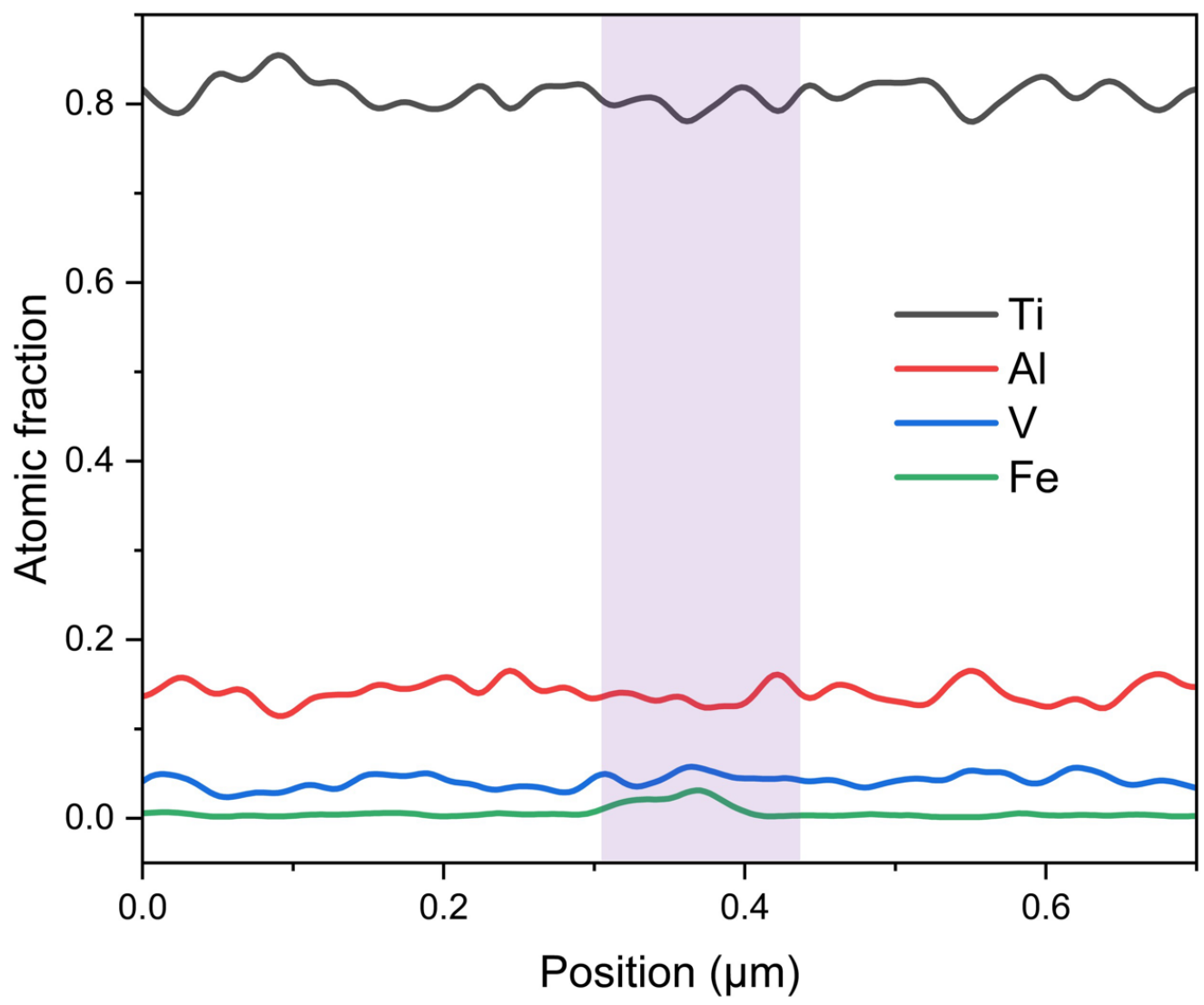

**Figure S6.** Compositional profiles of the as-cast $Ti_{81.4}Al_{16.8}V_{1.6}Fe_{0.2}$ alloy recorded across the yellow arrow in Figure 4e. The purple region indicates the Fe- and V-rich precipitate, and the result is consistent with the HAADF-STEM results in Figure 4e.

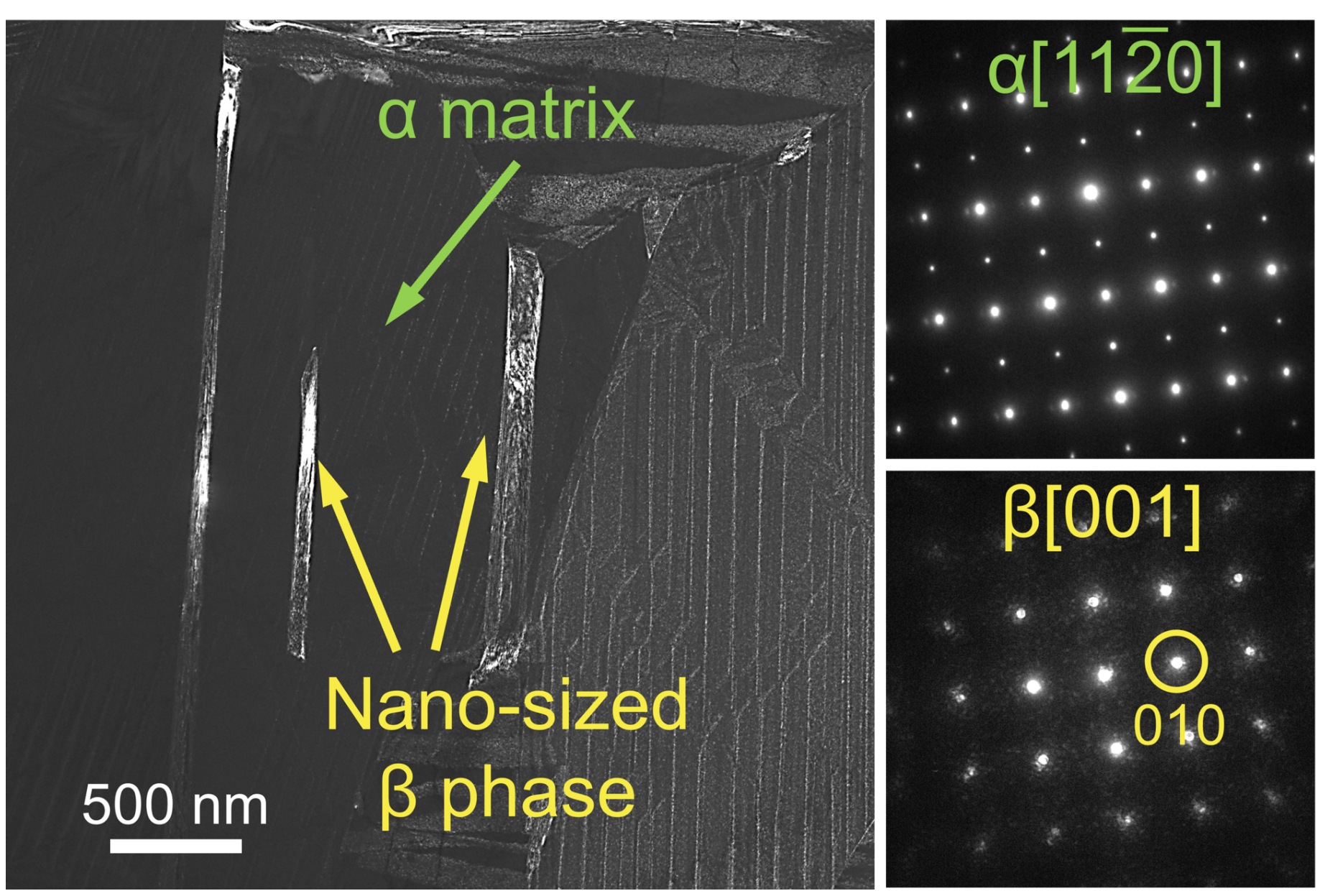

**Figure S7.** Dark-field TEM image of the as-cast $Ti_{81.4}Al_{16.8}V_{1.6}Fe_{0.2}$ alloy. The image was made from the (010) superlattice spot marked by the yellow circle showing the α matrix and lighted-up β phases. Right panels show Selected area electron diffraction (SAED) patterns from $[001]_{\beta}$ ZA and $[1120]_{\alpha}$ zone axes.

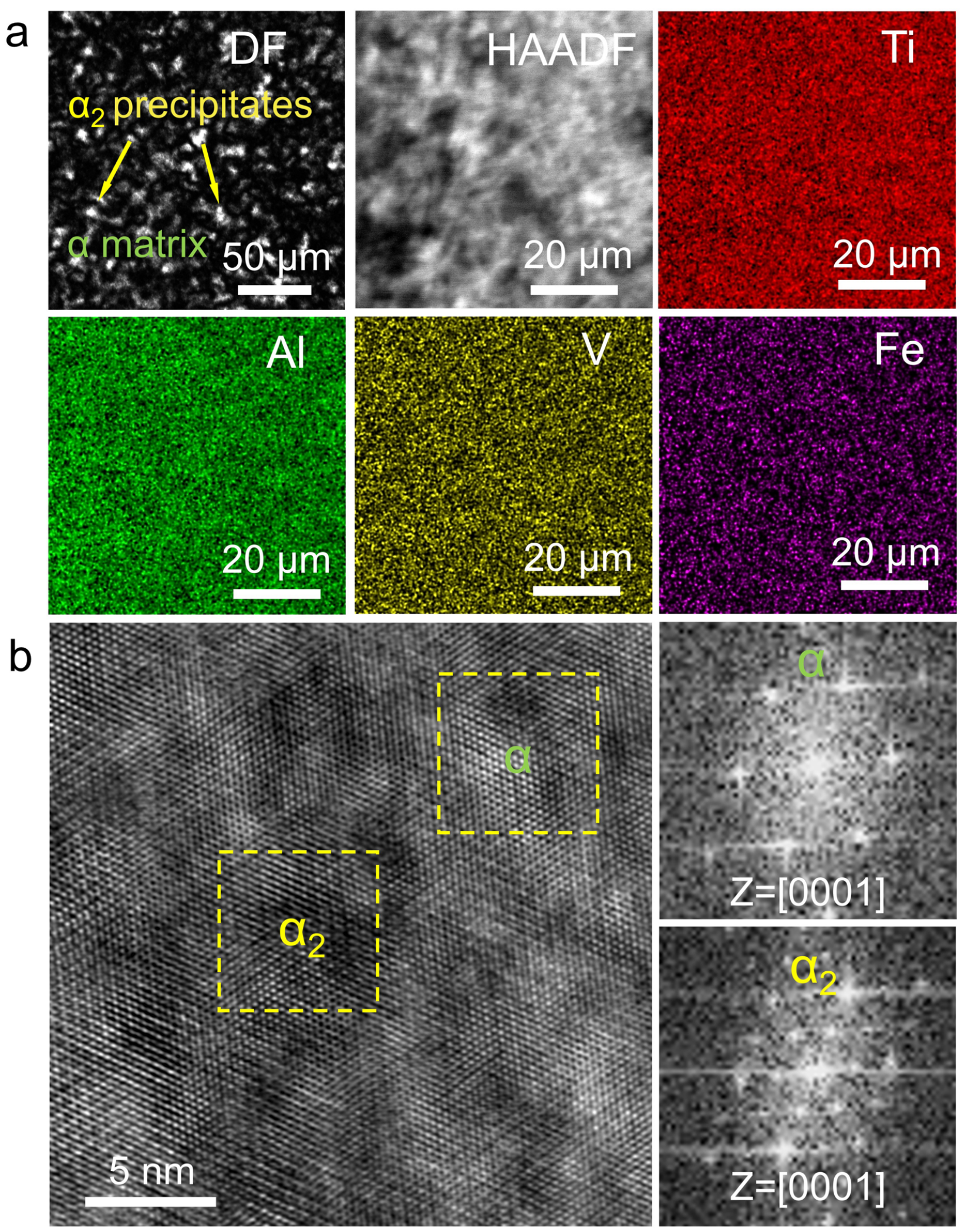

**Figure S8.** TEM images of the α-$\alpha_2$ region of the $Ti_{81.4}Al_{16.8}V_{1.6}Fe_{0.2}$ alloy. a) Dark-field TEM image, HAADF-STEM image, and EDS mappings of the α-$\alpha_2$ region. b) High-resolution TEM image of the α-$\alpha_2$ region, with their corresponding FFT images displayed on the right.

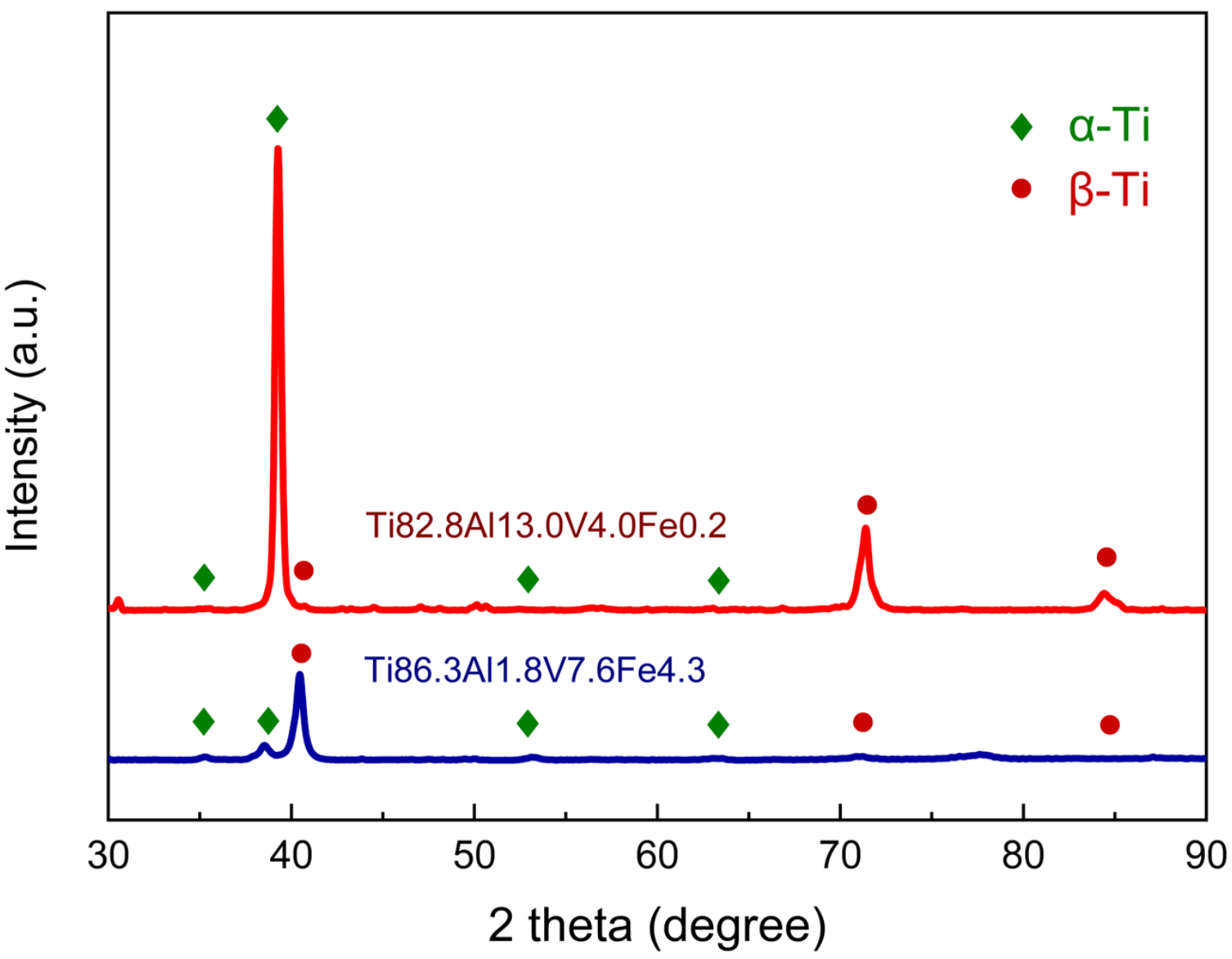

**Figure S9.** XRD patterns of the as-cast Ti-185 and $Ti_{82.8}Al_{13.0}V_{4.0}Fe_{0.2}$ alloys. Results show that the alloys were composed of α phase and β phase.

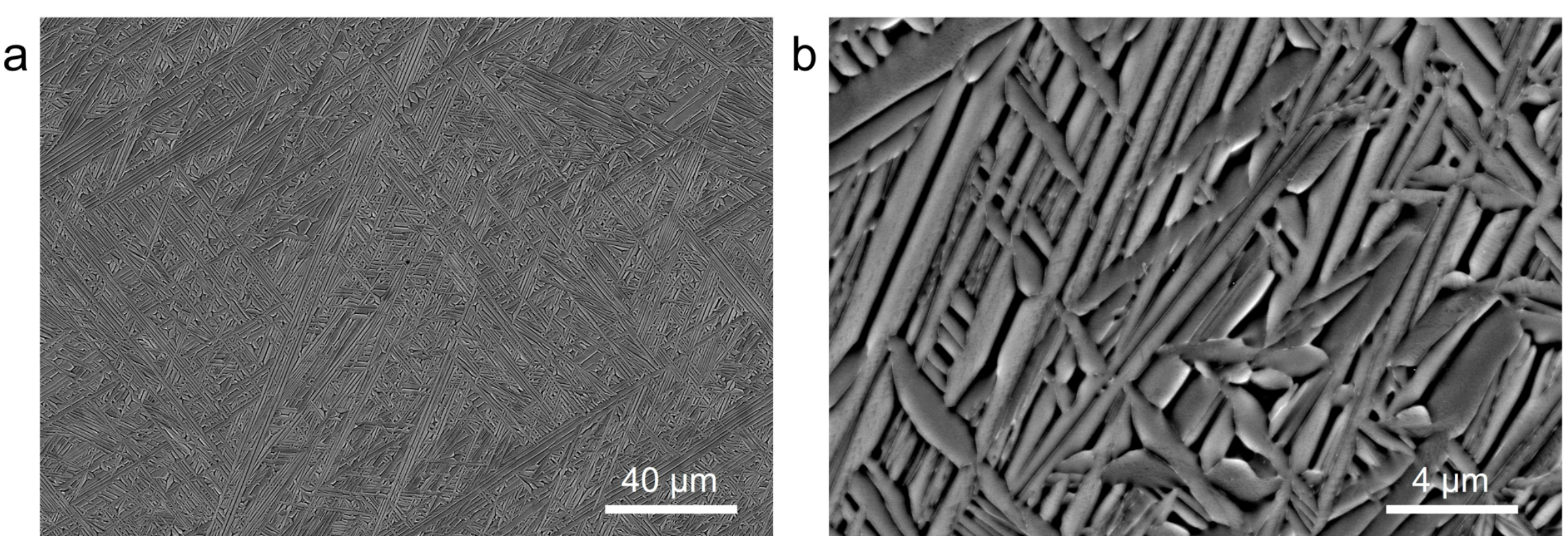

**Figure S10.** SEM images of $Ti_{82.8}Al_{13.0}V_{4.0}Fe_{0.2}$ alloy. a) At low magnification. b) At high magnification.

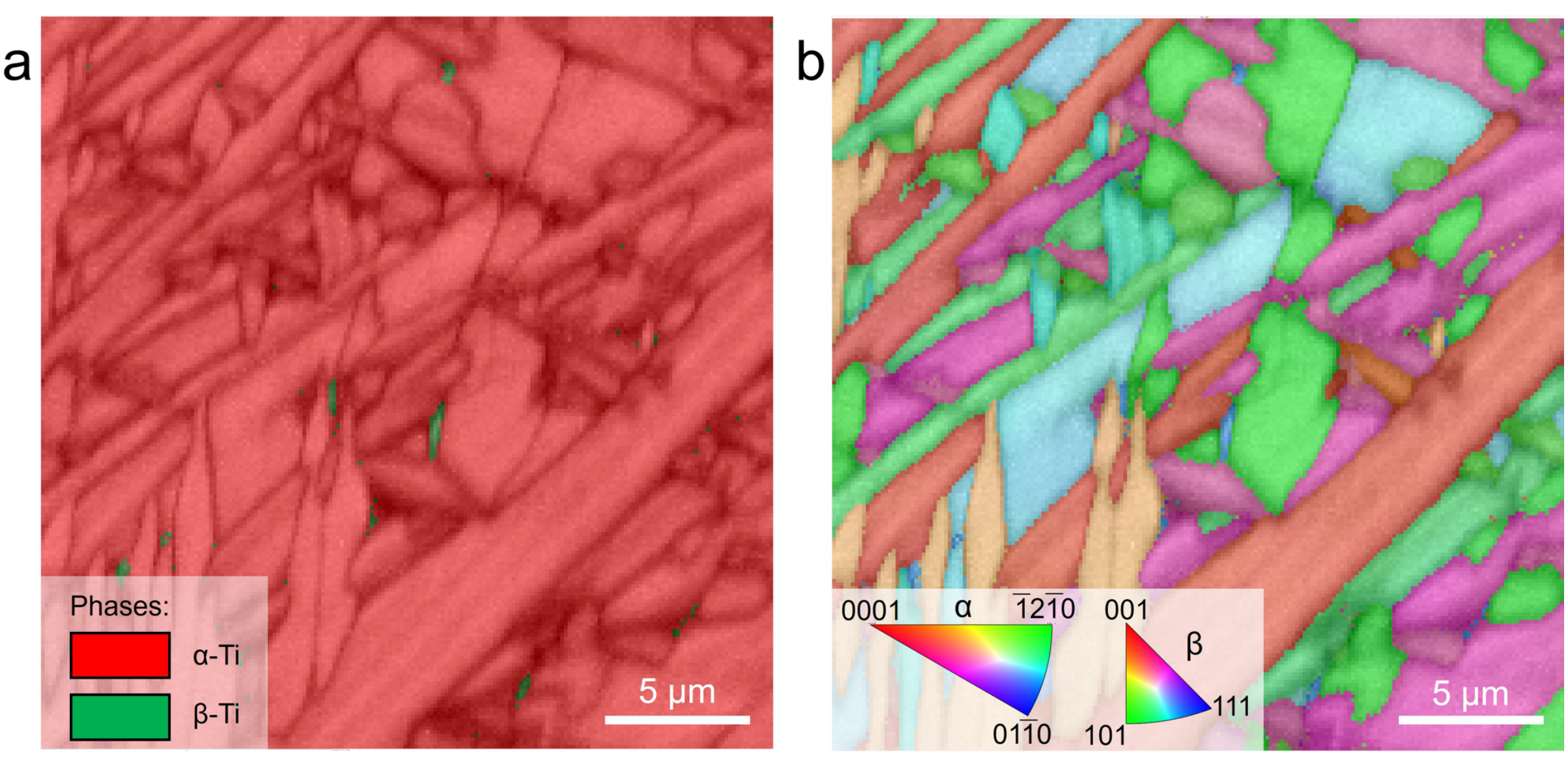

**Figure S11.** The EBSD images of the $Ti_{82.8}Al_{13.0}V_{4.0}Fe_{0.2}$ alloy. a) The phase map shows the alloy is composed of lath-like α-β structure. b) Its corresponding IPF map.

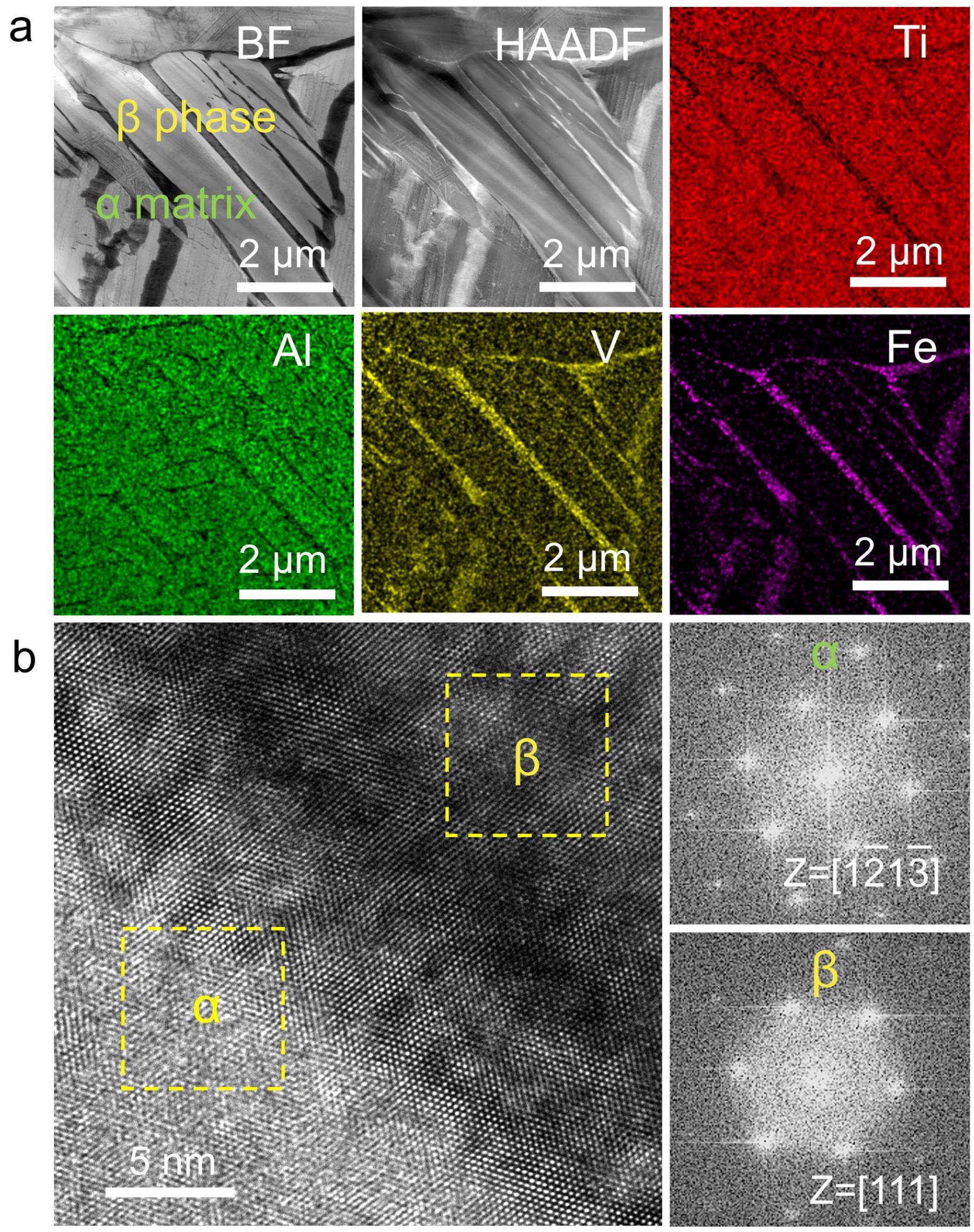

**Figure S12.** TEM images of the $Ti_{82.8}Al_{13.0}V_{4.0}Fe_{0.2}$ alloy. a) Bright-field TEM image, HAADF-STEM image, and EDS mappings of the α-β region. b) High-resolution TEM image of the α-β region, with their corresponding FFT images displayed on the right.

a

```
After composition_unify:
  Search Point: {'Co': 24.5, 'Al': 10.0, 'Cr': 23.0, 'Fe': 19.0, 'Ni': 23.5}
  Dependent Element: Co
  Composition: {'Al': 10.0, 'Cr': 23.0, 'Fe': 19.0, 'Ni': 23.5}
```

b

```
17:39:33,629 [pool-4-thread-1] INFO  CalculationEngine:   T=3273.15, P=100000, N=1, X(AL)=0.1, X(CR)=0.23, X(FE)=0.19, X(NI)=0.235
17:39:33,631 [pool-4-thread-1] INFO  CalculationEngine:   DEGREES OF FREEDOM 0
17:39:33,632 [pool-4-thread-1] INFO  CalculationEngine:   COMPONENT                STATUS    REF. STATE     T(K)          P(Pa)
17:39:33,632 [pool-4-thread-1] INFO  CalculationEngine:   VA                       ENTERED   SER
17:39:33,632 [pool-4-thread-1] INFO  CalculationEngine:   AL                       ENTERED   SER
17:39:33,632 [pool-4-thread-1] INFO  CalculationEngine:   CO                       ENTERED   SER
17:39:33,632 [pool-4-thread-1] INFO  CalculationEngine:   CR                       ENTERED   SER
17:39:33,632 [pool-4-thread-1] INFO  CalculationEngine:   FE                       ENTERED   SER
17:39:33,632 [pool-4-thread-1] INFO  CalculationEngine:   NI                       ENTERED   SER
17:39:37,788 [pool-4-thread-1] INFO  CalculationEngine:   You must release one of these conditions
17:39:37,789 [pool-4-thread-1] INFO  CalculationEngine:   T=3273.15, P=100000, N=1, X(AL)=0.1, X(CR)=0.23, X(FE)=0.19, X(NI)=0.235
17:39:37,789 [pool-4-thread-1] INFO  CalculationEngine:   DEGREES OF FREEDOM 0
17:39:47,849 [pool-4-thread-1] INFO  CalculationEngine:   PHASE CHANGE AT 1655.86850944
17:39:47,850 [pool-4-thread-1] INFO  CalculationEngine:  FCC_L12#1 forms
17:39:47,872 [pool-4-thread-1] INFO  CalculationEngine: Testing POLY result by global minimization procedure
17:39:48,093 [pool-4-thread-1] INFO  CalculationEngine: Calculated         45532 grid points in             0 s
17:39:48,731 [pool-4-thread-1] INFO  CalculationEngine:    CALCULATING SCHEIL SOLIDFICATION
17:39:48,813 [pool-4-thread-1] INFO  CalculationEngine:       T(C)      fraction solid
17:39:48,828 [pool-4-thread-1] INFO  CalculationEngine:     1382.809       0.000000
17:39:50,623 [pool-4-thread-1] INFO  CalculationEngine:   PHASE REGION:LIQUID + FCC_L12#1
17:39:50,624 [pool-4-thread-1] INFO  CalculationEngine:       T(C)      fraction solid
17:39:50,625 [pool-4-thread-1] INFO  CalculationEngine:     1382.713      0.2325986E-03
17:39:50,641 [pool-4-thread-1] INFO  CalculationEngine:     1381.713      0.4103561E-01
17:39:50,657 [pool-4-thread-1] INFO  CalculationEngine:     1380.713      0.7959357E-01
17:39:50,672 [pool-4-thread-1] INFO  CalculationEngine:     1379.713      0.1160581
17:39:50,689 [pool-4-thread-1] INFO  CalculationEngine:     1378.713      0.1505690
17:39:50,704 [pool-4-thread-1] INFO  CalculationEngine:     1377.713      0.1832548
```

c

```
17:39:55,108 [pool-4-thread-1] INFO  CalculationEngine:     1266.713      0.9285556
17:39:55,124 [pool-4-thread-1] INFO  CalculationEngine:     1265.713      0.9294523
17:39:55,138 [pool-4-thread-1] INFO  CalculationEngine:     1264.713      0.9303313
17:39:55,151 [pool-4-thread-1] INFO  CalculationEngine:     1263.713      0.9311931
17:39:55,430 [pool-4-thread-1] INFO  CalculationEngine:   Not true equilibrium, recalculating with global minimization
17:39:56,188 [pool-4-thread-1] INFO  CalculationEngine:   Not true equilibrium, recalculating with global minimization
17:39:56,875 [pool-4-thread-1] INFO  CalculationEngine:   Not true equilibrium, recalculating with global minimization
17:39:57,559 [pool-4-thread-1] INFO  CalculationEngine:   Not true equilibrium, recalculating with global minimization
17:39:58,231 [pool-4-thread-1] INFO  CalculationEngine:   Not true equilibrium, recalculating with global minimization
17:39:58,896 [pool-4-thread-1] INFO  CalculationEngine:   Not true equilibrium, recalculating with global minimization
17:39:59,290 [pool-4-thread-1] INFO  CalculationEngine:   PHASE REGION:BCC_B2#1 + FCC_L12#1
17:39:59,290 [pool-4-thread-1] INFO  CalculationEngine:       T(C)      fraction solid
17:39:59,291 [pool-4-thread-1] INFO  CalculationEngine:     1263.681       1.000000
17:39:59,316 [pool-4-thread-1] INFO  CalculationEngine:   Calculating properties ...
17:40:01,163 [pool-4-thread-1] INFO  CalculationEngine:  Liquidus temperature: 1655.869 K
17:40:01,163 [pool-4-thread-1] INFO  CalculationEngine:  Solidus temperature: 1536.831 K
17:40:01,209 [Thread-1] INFO  CalculationEngine:    POSTPROCESSOR VERSION 3.2
17:41:01,344 [Thread-1] ERROR CalculationEngine: Cleanup failed: TimeoutException. Thrown at: java.util.concurrent.FutureTask:get:LineNo:-1

After phase_volume_compute_with_scheil:
  Phase Volumes: {'FCC_L12': 0.9311931382863304, 'BCC_B2': 0.06880686171366956}
Press Enter to continue to the next step...
```

d

```
17:42:30,649 [Thread-1] INFO  CalculationEngine:  Creating a new composition set BCC_B2#3
17:42:30,652 [Thread-1] INFO  CalculationEngine: New CS: BCC_B2#3
17:42:30,652 [Thread-1] INFO  CalculationEngine: New CS: BCC_A2#3
17:43:31,629 [Thread-1] ERROR CalculationEngine: Cleanup failed: TimeoutException. Thrown at: java.util.concurrent.FutureTask:get:LineNo:-1

After tc_single_point:
  Mass: 53.49894499971497
  Volume: 6.90713077016198e-06
  Density: 7.745465777312967
Press Enter to continue to the next step...
```

e

```
17:45:03,451 [Thread-1] INFO  Py4JPropertyModelFactory: Creating Py4JPropertyModelFactory
17:45:11,003 [Thread-5] INFO  Py4JPropertyModel: 17:45:11,001 [Thread-1] INFO  CalculationEngine: Starting Thermo-Calc 2024b, Java-version: 21.0.2
17:45:11,775 [Thread-5] INFO  Py4JPropertyModel: 17:45:11,775 [Thread-1] INFO  SystemBuilder: Found matching previous system
17:45:13,683 [Thread-5] INFO  Py4JPropertyModel: 17:45:13,683 [Thread-1] INFO  JavaWrapper: *** Invoking Gibbs Energy System v6 ***
17:45:14,160 [Thread-5] INFO  Py4JPropertyModel: 17:45:14,159 [Thread-1] INFO  Py4JPropertyModelFactory: Creating Py4JPropertyModelFactory
17:45:16,620 [Thread-5] INFO  Py4JPropertyModel: 17:45:16,619 [pool-4-thread-1] INFO  CalculationEngine:   *** ERROR  2008 IN QMJBSV1: New composition sets needed
17:45:16,826 [Thread-5] INFO  Py4JPropertyModel: 17:45:16,826 [pool-6-thread-1] INFO  CalculationEngine:   *** ERROR  2008 IN QMJBSV1: New composition sets needed
17:45:16,931 [Thread-5] INFO  Py4JPropertyModel: 17:45:16,930 [pool-7-thread-1] INFO  CalculationEngine:  Creating a new composition set BCC_B2#3
17:45:16,933 [Thread-5] INFO  Py4JPropertyModel: 17:45:16,932 [pool-7-thread-1] INFO  CalculationEngine: New CS: BCC_B2#3
17:45:16,933 [Thread-5] INFO  Py4JPropertyModel: 17:45:16,933 [pool-7-thread-1] INFO  CalculationEngine: New CS: BCC_A2#3
17:45:17,590 [Thread-5] INFO  Py4JPropertyModel: 17:45:17,589 [pool-7-thread-1] INFO  CalculationEngine:  Creating a new composition set BCC_B2#4
17:45:17,590 [Thread-5] INFO  Py4JPropertyModel: 17:45:17,590 [pool-7-thread-1] INFO  CalculationEngine: New CS: BCC_B2#4
17:45:17,590 [Thread-5] INFO  Py4JPropertyModel: 17:45:17,590 [pool-7-thread-1] INFO  CalculationEngine: New CS: BCC_A2#4
17:45:17,757 [Thread-5] INFO  Py4JPropertyModel: User selected matrix phase: FCC_L12
17:46:59,770 [Thread-1] ERROR CalculationEngine: Cleanup failed: TimeoutException. Thrown at: java.util.concurrent.FutureTask:get:LineNo:-1

After get_total_yield_strength:
  Total Yield Strength (YS): 484.0474115013295
Press Enter to save the results and proceed to the next iteration...
```

**Figure S13.** Screenshots of the *Simulation Layer*'s workflow. To clearly illustrate the simulation logic, multithreading was disabled (i.e., one calculator node at a time), and execution was paused between each calculator invocation. a) Define system composition and dependent component as inputs. b) Launch Scheil Calculator to compute phase fractions. c) View phase-volume output from the Scheil calculation. d) Invoke TC-Single-Point Calculator to compute density. e) Run the Property Model Calculator to predict yield strength.

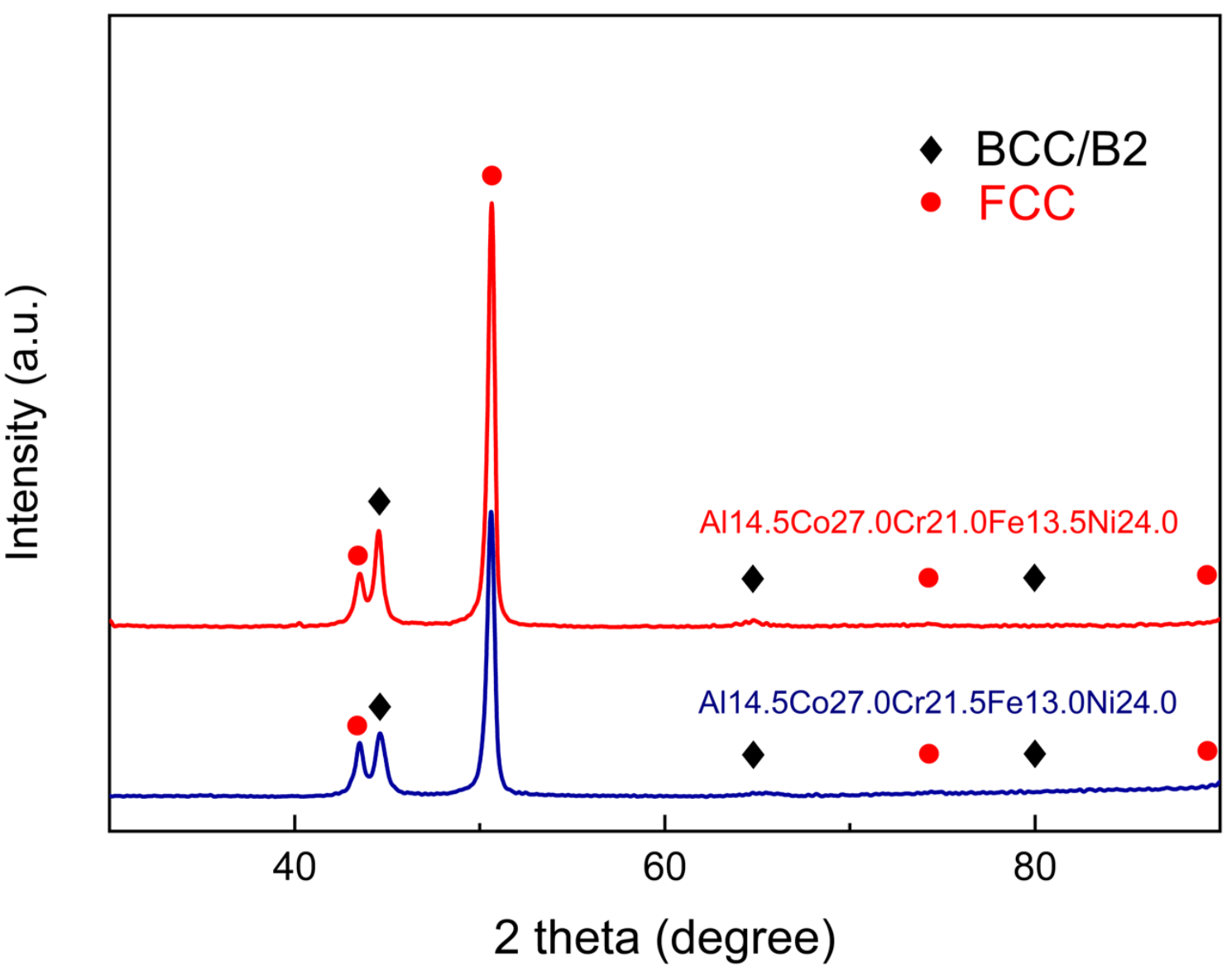

**Figure S14.** XRD patterns of the as-cast HEAs. Results showed that the alloys are composed of both FCC phase and BCC phase.

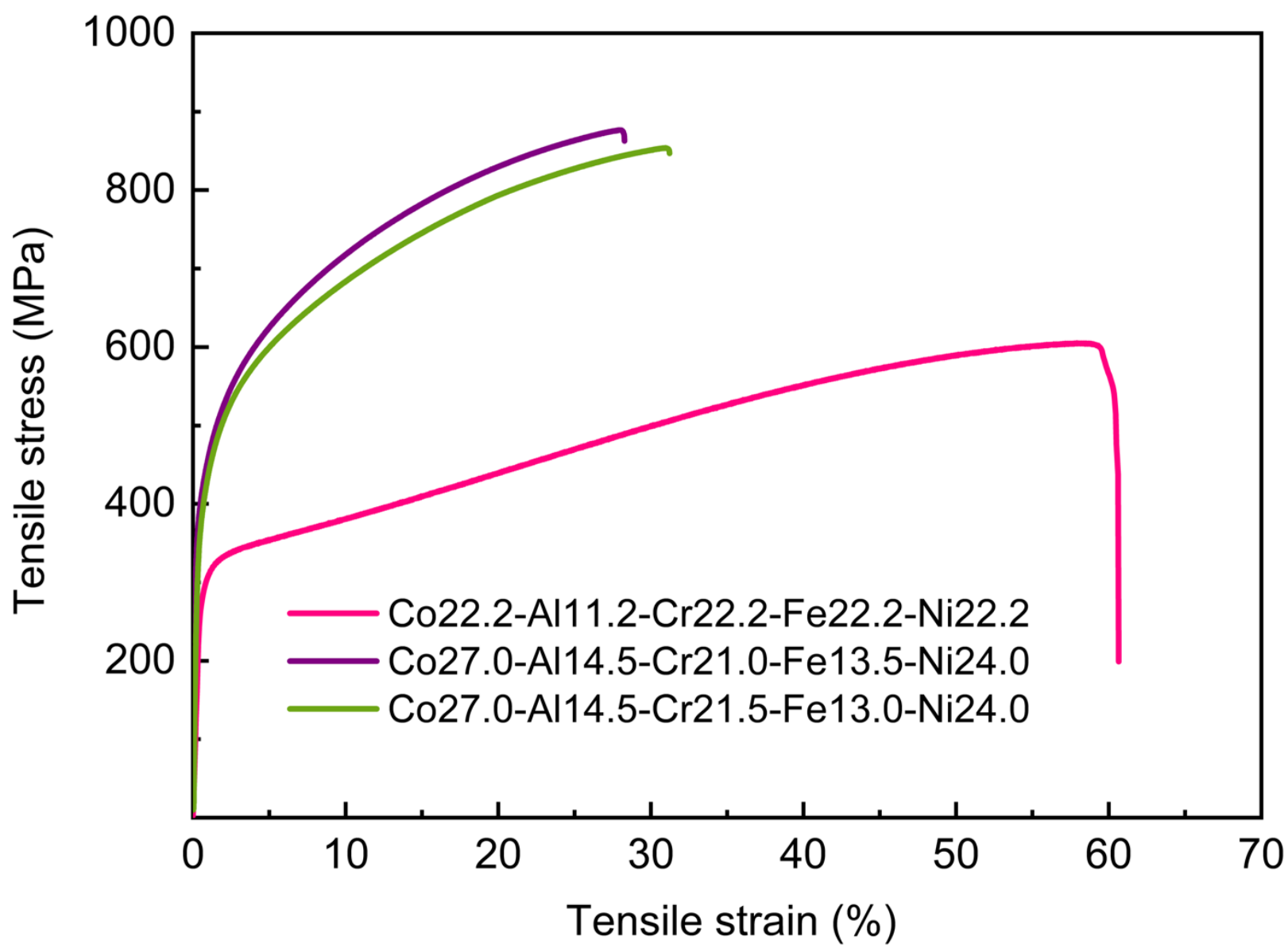

**Figure S15.** Typical engineering tensile stress-strain curves of the as-cast HEAs. The result confirmed the optimization of both strength and ductility after AutoMAT's prediction.

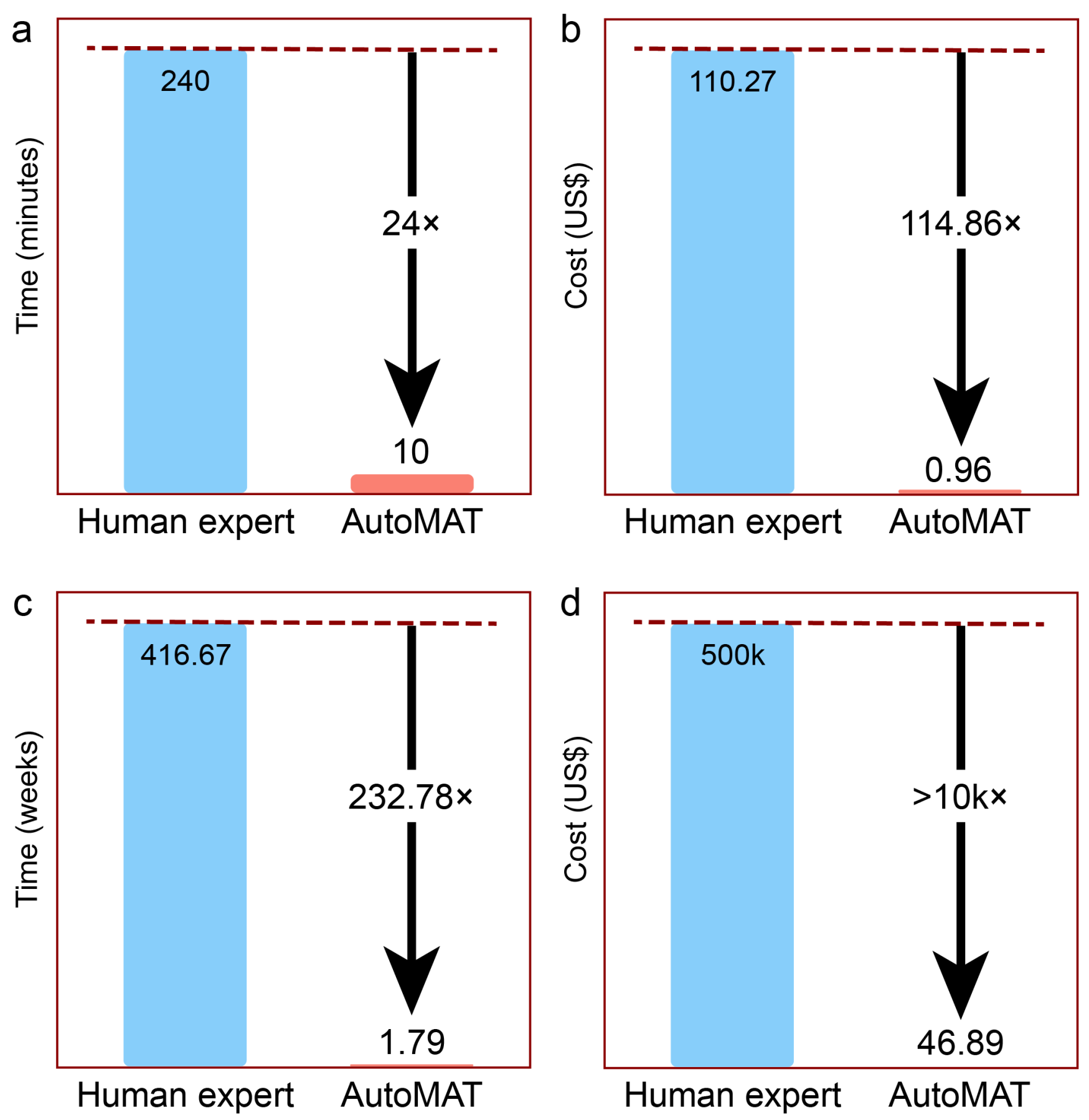

**Figure S16.** Time and cost savings achieved by AutoMAT's *Ideation* and *Simulation Layers*. a) Time comparison for the *Ideation Layer* in Case 1 showing a 24-fold reduction from 240 minutes (human expert) to 10 minutes (AutoMAT). b) Cost analysis for the *Ideation Layer* in Case 1 demonstrating 114.86-fold savings, from US$110.27 (human expert hourly rate based on US$5,000 monthly salary) to US$0.96 (AutoMAT operational cost). c) Time comparison for the *Simulation Layer* in Case 2 showing a 232.78-fold improvement, reducing task completion from 416.67 weeks (human expert) to 1.79 weeks (AutoMAT). d) Cost analysis for the *Simulation Layer* in Case 2 demonstrating over 10,000-fold savings, from approximately US$500,000 (human expert salary over extended timeframe) to US$46.89 (AutoMAT operational cost including 15-minute human setup time and computational resources at US$40 per two-week period in Singapore). Dashed lines indicate baseline human expert performance levels. All cost calculations include operational expenses and assume standard working hours for human experts.

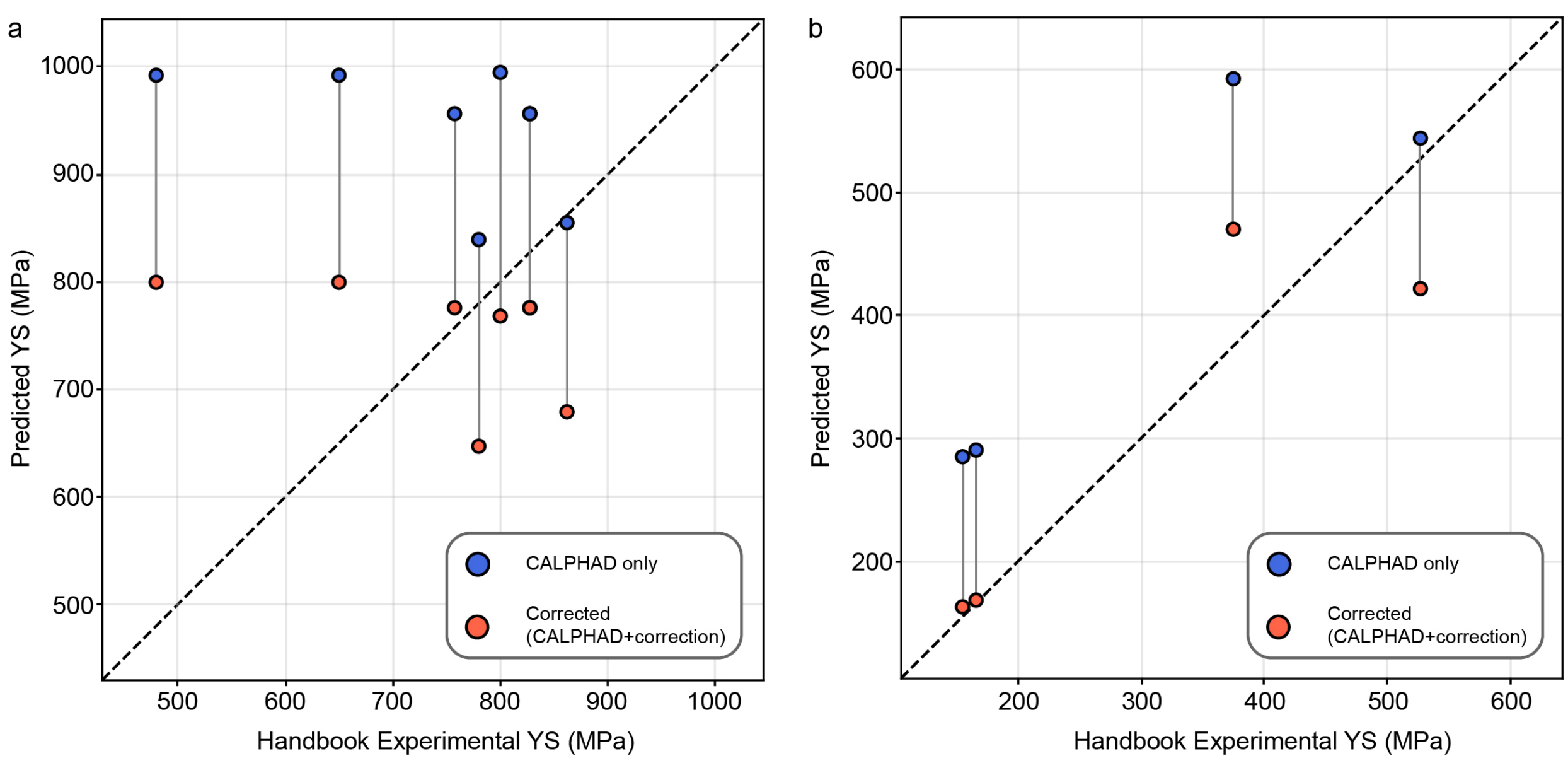

**Figure S17.** Performance of the data-driven yield strength correction on handbook training data. Predicted yield strength is plotted against handbook experimental values, where blue circles denote raw CALPHAD predictions, red circles denote corrected values (CALPHAD + Ridge regression residual), vertical lines connect the two predictions for each alloy, and the dashed line indicates parity. a) For titanium alloys, the raw CALPHAD predictions systematically overestimate experimental yield strength, and the correction shifts them toward the parity line. b) For HEAs, a similar overestimation is observed, and the correction again reduces the discrepancy between predicted and experimental values.

**Table S1.** Top 5 results from each iteration in Case 1.

| Iteration Number | Composition | $YS_{CALPHAD}$ | $YS_{corrected}$ | Density | Metric |
|---|---|---|---|---|---|
| 1 | $Ti_{86.0}Al_{10.0}V_{3.5}Fe_{0.5}$ | 961.42 | 776.53 | 4.460 | 8.979 |
| | $Ti_{85.5}Al_{10.0}V_{4.0}Fe_{0.5}$ | 961.11 | 774.76 | 4.466 | 8.904 |
| | $Ti_{86.0}Al_{9.5}V_{4.0}Fe_{0.5}$ | 964.70 | 776.76 | 4.471 | 8.883 |
| | $Ti_{86.0}Al_{10.0}V_{3.0}Fe_{1.0}$ | 956.91 | 769.36 | 4.468 | 8.825 |
| | $Ti_{85.0}Al_{10.0}V_{4.5}Fe_{0.5}$ | 960.79 | 772.98 | 4.473 | 8.822 |
| 2 | $Ti_{84.2}Al_{11.8}V_{3.8}Fe_{0.2}$ | 952.37 | 774.82 | 4.437 | 9.167 |
| | $Ti_{84.6}Al_{11.8}V_{3.0}Fe_{0.6}$ | 949.32 | 770.81 | 4.438 | 9.111 |
| | $Ti_{84.2}Al_{11.8}V_{3.6}Fe_{0.4}$ | 950.65 | 772.04 | 4.440 | 9.107 |
| | $Ti_{84.6}Al_{11.8}V_{2.8}Fe_{0.8}$ | 947.60 | 768.02 | 4.441 | 9.050 |
| | $Ti_{84.2}Al_{11.8}V_{3.4}Fe_{0.6}$ | 948.93 | 769.25 | 4.438 | 9.092 |
| 3 | $Ti_{82.8}Al_{13.0}V_{4.0}Fe_{0.2}$ | 944.88 | 770.58 | 4.427 | 9.209 |
| | $Ti_{83.2}Al_{13.2}V_{3.0}Fe_{0.6}$ | 941.13 | 767.08 | 4.423 | 9.204 |
| | $Ti_{82.2}Al_{13.4}V_{4.2}Fe_{0.2}$ | 942.23 | 768.62 | 4.425 | 9.204 |
| | $Ti_{83.4}Al_{13.0}V_{3.0}Fe_{0.6}$ | 942.28 | 767.60 | 4.425 | 9.191 |
| | $Ti_{83.2}Al_{13.4}V_{2.6}Fe_{0.8}$ | 938.59 | 764.70 | 4.421 | 9.193 |
| 4 | $Ti_{82.2}Al_{15.0}V_{2.6}Fe_{0.2}$ | 935.35 | 771.51 | 4.387 | 9.596 |
| | $Ti_{81.6}Al_{15.4}V_{2.8}Fe_{0.2}$ | 932.76 | 769.61 | 4.386 | 9.582 |
| | $Ti_{82.8}Al_{15.0}V_{1.6}Fe_{0.6}$ | 933.12 | 768.9 | 4.386 | 9.573 |
| | $Ti_{82.2}Al_{15.4}V_{1.8}Fe_{0.6}$ | 930.63 | 767.10 | 4.384 | 9.570 |
| | $Ti_{83.0}Al_{14.8}V_{1.6}Fe_{0.6}$ | 934.20 | 769.34 | 4.388 | 9.560 |
| 5 | $Ti_{81.4}Al_{16.8}V_{1.6}Fe_{0.2}$ | 927.08 | 771.90 | 4.355 | 9.913 |
| | $Ti_{82.0}Al_{16.8}V_{0.6}Fe_{0.6}$ | 925.19 | 769.63 | 4.353 | 9.904 |
| | $Ti_{82.2}Al_{16.6}V_{0.6}Fe_{0.6}$ | 926.26 | 770.06 | 4.355 | 9.889 |
| | $Ti_{81.2}Al_{16.8}V_{1.8}Fe_{0.2}$ | 926.69 | 770.93 | 4.357 | 9.881 |
| | $Ti_{81.8}Al_{16.8}V_{0.8}Fe_{0.6}$ | 924.81 | 768.67 | 4.356 | 9.862 |

**Table S2.** Chemical compositions of the as-cast titanium alloys (in wt %).

| Alloy | Composition | Ti (%) | Al (%) | V (%) | Fe (%) |
|---|---|---|---|---|---|
| $Ti_{86.3}Al_{1.8}V_{7.6}Fe_{4.3}$ | Nominal | 86.00 | 1.00 | 8.00 | 5.00 |
| | Experimental | 86.09 | 1.08 | 7.93 | 4.90 |
| $Ti_{82.8}Al_{13.0}V_{4.0}Fe_{0.2}$ | Nominal | 87.51 | 7.74 | 4.50 | 0.25 |
| | Experimental | 87.62 | 7.67 | 4.44 | 0.27 |
| $Ti_{81.4}Al_{16.8}V_{1.6}Fe_{0.2}$ | Nominal | 87.71 | 10.20 | 1.83 | 0.26 |
| | Experimental | 87.30 | 10.61 | 1.82 | 0.27 |

**Table S3.** Tensile properties of the as-cast titanium alloys.

| Alloy | Density ($g/cm^3$) | Yield strength (MPa) | Ultimate tensile strength (MPa) | Engineering strain (%) |
|---|---|---|---|---|
| $Ti_{86.3}Al_{1.8}V_{7.6}Fe_{4.3}$ | 4.70 | 831±6 | 872±6 | 7.0±0.3 |
| $Ti_{82.8}Al_{13.0}V_{4.0}Fe_{0.2}$ | 4.40 | 865±5 | 984±13 | 4.3±0.2 |
| $Ti_{81.4}Al_{16.8}V_{1.6}Fe_{0.2}$ | 4.32 | 943±17 | 1053±20 | 3.6±0.4 |

**Table S4.** Theoretical and experimental yield strength of the alloys in Figure 4c. The yield strength is calculated at 298 K (equilibrium calculation); Grain size is set to 100 micrometers.

| Alloy | Theoretical yield strength (MPa) | Database | Experimental yield strength (MPa) | Reference |
|---|---|---|---|---|
| Mg-Y11-Al1 (wt%) | 92 | TCMG8 | 350 | [1] |
| Al-Mg6.0-Sc0.3 (wt%) | 63 | TCAL10 | 200 | [2] |
| Ti96.63-0.14O-3.23Fe (wt%) | 628 | TCTI6 | 744 | [3] |
| Ti75.25-Al20-Cr4.75 (at%) | N/A | TCTI6 | 800 | [4] |
| Ti82-5Al-5Mo-5V-3Cr (wt%) | 551 | TCTI6 | 830 | [5] |
| Fe41.2-Mn35.2-Co11.8-Cr11.8 (at%) | 194 | TCHEA8 | 200 | [6] |
| Al2.4Co24.4-Cr24.4-Fe24.4-Ni24.4 (at%) | 347 | TCHEA8 | 185 | [7] |
| Nb45-Ta25-Ti15-Hf15 (at%) | 472 | TCHEA8 | 596 | [8] |
| Al19-Fe20-Co20-Ni41 (at%) | 442 | TCHEA8 | 670 | [9] |
| Ti25-Zr25-Hf25-Nb25 (at%) | 371 | TCHEA8 | 750 | [10] |
| Cr20-Mn20-Fe20-Co20-Ni20 (at%) | 136 | TCHEA8 | 410 | [11] |

**Table S5.** The EDS chemical compositions of the α phase and β phase of the as-cast $Ti_{81.4}Al_{16.8}V_{1.6}Fe_{0.2}$ alloy (in at%).

| Phases | Ti (%) | Al (%) | V (%) | Fe (%) |
|---|---|---|---|---|
| α | 83.88±2.30 | 13.97±1.99 | 1.88±0.39 | 0.27±0.07 |
| β | 79.81±2.81 | 12.65±1.73 | 4.62±0.88 | 2.92±0.56 |

**Table S6.** Theoretical crack susceptibility coefficient (CSC) and volumetric shrinkage of the as-cast titanium alloys and HEAs. The CSC is calculated via the Easton model and the calculation is based on the classic Scheil non-equilibrium solidification simulation.

| Alloy | CSC (Easton) | Volumetric shrinkage (vol.%) | Database used |
|---|---|---|---|
| Ti-185 | 333.15960 | 8.66584 | TCTI6 |
| Ti-10Al-3.5V-0.5Fe | 363.85259 | 6.94958 | TCTI6 |
| Ti-11.8Al-3.8V-0.2Fe | 361.84301 | 6.24603 | TCTI6 |
| Ti-13Al-4V-0.2Fe | 363.59240 | 5.76828 | TCTI6 |
| Ti-15Al-2.6V-0.2Fe | 375.92292 | 7.49232 | TCTI6 |
| Ti-16.8Al-1.6V-0.2Fe | 385.59387 | 8.95047 | TCTI6 |
| Co22.2-Al11.2-Cr22.2-Fe22.2-Ni22.2 | 45.33669 | 12.42480 | TCHEA7 |
| Co23.0-Al13.5-Cr23-Fe19.5-Ni21 | 19.43483 | 13.02764 | TCHEA7 |
| Co24-Al13.5-Cr24-Fe17.5-Ni21 | 21.22217 | 13.23625 | TCHEA7 |
| Co25.5-Al14-Cr22.5-Fe15.5-Ni22.5 | 15.61612 | 13.59171 | TCHEA7 |
| Co27-Al14.5-Cr21-Fe13.5-Ni24 | 11.00133 | 14.00847 | TCHEA7 |
| Co27-Al14.5-Cr21.5-Fe13-Ni24 | 11.47189 | 14.05023 | TCHEA7 |

**Table S7.** The EDS chemical compositions of the α phase and β phase of the $Ti_{82.8}Al_{13.0}V_{4.0}Fe_{0.2}$ alloy (in at%).

| Phases | Ti (%) | Al (%) | V (%) | Fe (%) |
|---|---|---|---|---|
| α | 84.52±2.28 | 11.30±1.63 | 3.99±0.79 | 0.19±0.08 |
| β | 76.35±3.41 | 7.03±1.03 | 13.43±2.35 | 3.19±0.63 |

**Table S8.** Top 5 results from each iteration in Case 2.

| Iteration Number | Composition | $YS_{CALPHAD}$ (MPa) | $YS_{CALPHAD}$ (MPa) |
|---|---|---|---|
| | $Al_{13.5}Co_{23.0}Cr_{23.0}Fe_{19.5}Ni_{21.0}$ | 810.58 | 688.26 |
| | $Al_{13.5}Co_{23.0}Cr_{22.5}Fe_{20.0}Ni_{21.0}$ | 806.84 | 684.52 |
| 1 | $Al_{13.5}Co_{23.0}Cr_{21.5}Fe_{21.5}Ni_{20.5}$ | 802.93 | 680.61 |
| | $Al_{13.5}Co_{23.0}Cr_{21.0}Fe_{22.0}Ni_{20.5}$ | 799.55 | 677.23 |
| | $Al_{13.5}Co_{23.0}Cr_{20.5}Fe_{22.5}Ni_{20.5}$ | 796.32 | 674.00 |
| | $Al_{13.5}Co_{24.0}Cr_{24.0}Fe_{17.5}Ni_{21.0}$ | 851.57 | 729.25 |
| | $Al_{14.0}Co_{24.5}Cr_{21.0}Fe_{18.0}Ni_{22.5}$ | 850.60 | 728.26 |
| 2 | $Al_{13.5}Co_{24.5}Cr_{23.0}Fe_{18.5}Ni_{20.5}$ | 849.38 | 727.06 |
| | $Al_{13.5}Co_{24.0}Cr_{23.5}Fe_{18.0}Ni_{21.0}$ | 847.84 | 725.52 |
| | $Al_{13.5}Co_{24.5}Cr_{22.5}Fe_{19.0}Ni_{20.5}$ | 847.31 | 724.99 |
| | $Al_{14.0}Co_{25.5}Cr_{22.5}Fe_{15.5}Ni_{22.5}$ | 886.70 | 764.36 |
| | $Al_{14.0}Co_{25.5}Cr_{22.0}Fe_{16.0}Ni_{22.5}$ | 886.50 | 764.16 |
| 3 | $Al_{14.0}Co_{25.0}Cr_{22.5}Fe_{16.0}Ni_{22.5}$ | 877.66 | 755.32 |
| | $Al_{14.0}Co_{25.0}Cr_{22.0}Fe_{16.5}Ni_{22.5}$ | 874.25 | 751.91 |
| | $Al_{13.5}Co_{25.0}Cr_{25.0}Fe_{15.5}Ni_{21.0}$ | 865.95 | 743.63 |
| | $Al_{14.5}Co_{27.0}Cr_{21.0}Fe_{13.5}Ni_{24.0}$ | 904.90 | 782.55 |
| | $Al_{14.5}Co_{27.0}Cr_{20.5}Fe_{14.0}Ni_{24.0}$ | 902.97 | 780.62 |
| 4 | $Al_{14.0}Co_{27.0}Cr_{21.5}Fe_{15.5}Ni_{22.0}$ | 895.16 | 772.82 |
| | $Al_{14.0}Co_{27.0}Cr_{21.0}Fe_{16.0}Ni_{22.0}$ | 895.10 | 772.76 |
| | $Al_{14.0}Co_{27.0}Cr_{22.0}Fe_{15.0}Ni_{22.0}$ | 895.02 | 772.68 |
| | $Al_{14.5}Co_{27.0}Cr_{21.5}Fe_{13.0}Ni_{24.0}$ | 906.64 | 784.29 |
| | $Al_{14.5}Co_{27.0}Cr_{22.0}Fe_{12.5}Ni_{24.0}$ | 906.09 | 783.74 |
| 5 | $Al_{14.5}Co_{28.0}Cr_{20.5}Fe_{13.5}Ni_{23.5}$ | 905.32 | 782.97 |
| | $Al_{14.5}Co_{26.0}Cr_{22.5}Fe_{12.5}Ni_{24.5}$ | 905.00 | 782.65 |
| | $Al_{14.5}Co_{26.0}Cr_{22.0}Fe_{13.0}Ni_{24.5}$ | 904.75 | 782.40 |

**Table S9.** Chemical compositions of the as-cast HEAs (in wt %).

| Alloy | Composition | Al (%) | Co (%) | Cr (%) | Fe (%) | Ni (%) |
|---|---|---|---|---|---|---|
| $Al_{14.5}Co_{27.0}Cr_{21.5}Fe_{13.0}Ni_{24.0}$ | Nominal | 7.47 | 30.40 | 21.35 | 13.87 | 26.91 |
| | Experimental | 7.51 | 30.36 | 21.42 | 13.86 | 26.85 |
| $Al_{14.5}Co_{27.0}Cr_{21.0}Fe_{13.5}Ni_{24.0}$ | Nominal | 7.47 | 30.38 | 20.85 | 14.40 | 26.90 |
| | Experimental | 7.48 | 30.33 | 20.88 | 14.39 | 26.92 |

**Table S10.** Tensile properties of the as-cast HEAs.

| Alloy | Yield strength (MPa) | Ultimate tensile strength (MPa) | Engineering strain (%) |
|---|---|---|---|
| $Al_{11.2}Co_{22.2}Cr_{22.2}Fe_{22.2}Ni_{22.2}$ | 305±9 | 618±13 | 56.2±5.1 |
| $Al_{14.5}Co_{27.0}Cr_{21.5}Fe_{13.0}Ni_{24.0}$ | 378±10 | 854±24 | 25.2±6.0 |
| $Al_{14.5}Co_{27.0}Cr_{21.0}Fe_{13.5}Ni_{24.0}$ | 397±6 | 857±20 | 26.0±2.3 |